\documentclass[prd,superscriptaddress,showpacs,amssymb,amsmath,amsfonts,aps,altaffilletter,nofootinbib,letterpaper,twocolumn]{revtex4-1}

\usepackage{amssymb}
\usepackage{color}
\usepackage{graphicx}
\usepackage{float}
\usepackage{caption}
\usepackage{subcaption}
\usepackage{multirow}
\usepackage{array}
\usepackage{hyperref}
\usepackage{bm}
\usepackage[normalem]{ulem}

\newcommand{\Msun}{\mathrm{M}_\odot}

\newcommand{\ip}[2]{\left< #1, #2 \right>}

\newcommand{\signal}{\ensuremath{h^{\dagger}}}
\newcommand{\template}{\ensuremath{h}}
\newcommand{\rhonew}{\tilde{\rho}}
\newcommand{\rstat}{\hat{\rho}}
\newcommand{\threshrstat}{\rstat^\dagger}
\newcommand{\rhoopt}{\rho_{\textrm{opt}}}
\newcommand{\eff}{\mathcal{E}}

\newcommand{\vol}{\mathcal{V}}

\newcommand{\intparams}{\mathbf{\Upsilon}}
\newcommand{\extparams}{\mathbf{\Xi}}
\newcommand{\allparams}{\mathbf{\Lambda}}
\newcommand{\spatialV}{\mathbf{x}}
\newcommand{\fap}{\mathcal{F}}

\newcommand{\exTileRecoloredGain}{\ensuremath{10\pm6}}

\graphicspath{{figures/}{figures/effectualness/}}

\begin{document}

\title{Implementing a search for gravitational waves from non-precessing, spinning binary black holes}

\author{Collin Capano} 

\affiliation{Max Planck Institute for Gravitational Physics (Albert Einstein Institute), Callinstrasse 38, 30167 Hannover, Germany} 

\author{Ian Harry} 

\affiliation{Max Planck Institute for Gravitational Physics (Albert Einstein Institute), Am M\"uhlenberg 1, 14476 Potsdam, Germany} 

\author{Stephen Privitera}

\affiliation{Max Planck Institute for Gravitational Physics (Albert Einstein Institute), Am M\"uhlenberg 1, 14476 Potsdam, Germany} 

\author{Alessandra Buonanno} 

\affiliation{Max Planck Institute for Gravitational Physics (Albert Einstein Institute), Am M\"uhlenberg 1, 14476 Potsdam, Germany} 
\affiliation{Department of Physics, University of Maryland, College Park, MD 20742, USA}

\begin{abstract}

Searching for gravitational waves (GWs) from binary black holes (BBHs) with
LIGO and Virgo involves matched-filtering data against a set of representative
signal waveforms --- a template bank --- chosen to cover the full signal space
of interest with as few template waveforms as possible. Although the component
black holes may have significant angular momenta (spin), previous searches for
BBHs have filtered LIGO and Virgo data using only waveforms where both
component spins are zero. This leads to a loss of signal-to-noise ratio for
signals where this is not the case.  Combining the best available template
placement techniques and waveform models, we construct a template bank of GW
signals from BBHs with component spins $\chi_{1,2}\in [-0.99, 0.99]$ aligned
with the orbital angular momentum, component masses $m_{1,2}\in [2,
48]\,\mathrm{M}_\odot$, and total mass $M_\mathrm{total} \leq
50\,\mathrm{M}_\odot$. Using effective-one-body waveforms with spin effects, we
show that less than $3\%$ of the maximum signal-to-noise ratio (SNR) of these
signals is lost due to the discreetness of the bank, using the early advanced
LIGO noise curve. We use simulated advanced LIGO noise to compare the
sensitivity of this bank to a non-spinning bank covering the same parameter
space. In doing so, we consider the competing effects between improved SNR and
signal-based vetoes, and the increase in the rate of false alarms of the
aligned-spin bank due to covering a larger parameter space. We find that the
aligned-spin bank can be a factor of $1.3$ -- $5$ more sensitive than a
non-spinning bank to BBHs with dimensionless spins $> +0.6$ and component
masses $\gtrsim 20\,\mathrm{M}_\odot$. Even larger gains are obtained for systems
with equally high spins but smaller component masses.

\end{abstract}

\maketitle

\section{Introduction}

On September 14, 2015 the advanced LIGO (aLIGO) detectors made the first
detection of gravitational waves from two coalescing black holes,
GW150914~\cite{Abbott:2016blz}. Two types of searches found this event: an
unmodeled search designed to look for coherent ``bursts'' of power in both
LIGO detectors~\cite{TheLIGOScientific:2016uux} and a modeled search, designed
to search for gravitational waves (GWs) from coalescing binary neutron stars
(BNS), neutron-star black hole binaries (NSBH), and binary black holes
(BBH)~\cite{TheLIGOScientific:2016qqj}. The primary difference between these
searches is that the modeled search uses a bank of template waveforms of
expected signals to match filter the data, obtaining a signal-to-noise ratio
(SNR) for candidate events.  Using the morphology of the template, additional
signal-based vetoes are applied to better separate candidate events from
background noise \cite{Allen:2004gu}. This can lead to a more sensitive search
than what can be obtained from unmodeled searches.  For instance, the modeled
search found GW150914 with significance $>5.1\,\sigma$, while in the unmodeled
search the event had a significance of $4.4\,\sigma$ \cite{Abbott:2016blz}.

Key to maximizing the sensitivity of the modeled search is that the parameters
of the template waveforms are sufficiently close to sources' parameters such
that the morphology of waveforms matches that of signals. Any mismatch between
signal and template leads to a loss in SNR, and down-weighting by the
signal-based vetoes. Some source parameters, such as the coalescence time and
phase, can be analytically maximized over, resulting in essentially no SNR
loss.  The remaining parameters, however, are traditionally covered by some
gridding of the parameter space, in which a small but non-zero amount of SNR
is lost to signals from systems not lying exactly on the grid.

Sufficiently covering the space of possible signals is challenging,
particularly for BBHs, for which there is a large range of possible masses and
spins. These systems are thought to form predominantly through the
co-evolution of massive ($\gtrsim 15\,\Msun$) stars in \emph{field}
binaries~\cite{Belczynski:2011bv, Dominik:2012kk, Ferdman:2014rna,
Bulik:2008ab}, or by the dynamical capture of two independently formed black
holes (BHs) living in dense stellar environments, such as globular clusters or
galactic cores~\cite{Joshi:1999vf, Fregeau:2004if, Pooley:2003zb,
Ivanova:2007bu, Clausen:2012zu, Belczynski:2014iua, Rodriguez:2015oxa}.  Prior
to the discovery of GW150914, stellar-mass BHs were known from observations of
X-ray binaries, in which a black hole accretes matter from a stellar
companion.  The mass of the black hole in these binaries, most of which are in
the galaxy, are between $5$ -- $20\,\Msun$. The BHs in two extra galactic
binaries, IC10 X-1 and NGC300 X-1, were estimated to have masses between
$21$-$35\,\Msun$ \cite{Prestwich:2007mj, Silverman:2008ss} and
$12$-$24\,\Msun$ \cite{Crowther:2010}, respectively, but these estimates were
later questioned~\cite{Laycock:2015qla}. Even so, population synthesis models
have predicted that isolated BHs formed from stellar evolution could have
masses exceeding $30\,\Msun$ depending largely on the metallicity of the
progenitor stars \cite{Dominik:2012kk, Belczynski:2009xy}. This prediction was
confirmed with the detection of GW150914: its component masses were determined
to be $36^{+5}_{-4}\,\Msun$ and
$29^{+4}_{-4}\,\Msun$~\cite{TheLIGOScientific:2016wfe}.

The mutual interactions between the two components as they evolve and collapse
into BHs in a field binary affects their masses. Population synthesis
models predict BBHs with total masses up to $\sim 100\,\Msun$
\cite{Dominik:2013tma}.  It may be possible to form even more massive binaries
from the dynamical capture of two BHs that formed independently in a low (less
than solar) metallicity environment.  GW150914 was consistent with both of
these formation channels \cite{TheLIGOScientific:2016htt}.
This leaves a large parameter space for BBH searches to cover. Previous
searches have cast as wide a net as possible, probing for GWs from
BBHs as light as $M_\mathrm{total} = 0.4M_\odot$~\cite{Abbott:2005pf,
  Abbott:2007xi} and as heavy as $M_\mathrm{total} = 400
\Msun$~\cite{Aasi:2014bqj, Mohapatra:2014rda}, above which the
signal is entirely out of band.

Efficient coverings of the mass parameter space have been available
for quite some time~\cite{Owen:1998dk, Cokelaer:2007kx}.  These
consist of two basic flavors: lattice-based~\cite{Brown:2012qf,
  Harry:2013tca} and stochastic-based~\cite{Babak:2008rb,
  Harry:2009ea, Ajith:2012mn, Privitera:2013xza}. Lattice-based
techniques are most profitably applied to low-mass systems, for which
the merger and ringdown occur out of band and only the inspiral
portion of the waveform contributes to the SNR. In this case, one can
construct a special set of coordinates in which a regular lattice is
the optimal placement strategy. Stochastic-based techniques, by
contrast, are completely generic, but are not guaranteed to be optimal
and quickly become computationally limited as the required template
bank size increases, as with increasing parameter space dimension or
improved detector bandwidth.

A substantially harder problem is how to cover the remaining six-dimensional
spin parameter space for BBHs, where lattice-based techniques do not directly
apply, and stochastic-based techniques approach computational limitations.
However, building a template bank with spin effects may be crucial to
optimizing the detection rate in these searches. Electromagnetic observations
of BHs in X-ray binaries ~\cite{McClintock:2006xd, Gou:2011nq, Gou:2010qq,
Gou:2009ks, Miller:2009cw, Valtonen:2010xf, Martocchia:2002wt}, as well as
population synthesis models for BBH formation~\cite{Belczynski:2007xg},
indicate the potential for a range of BH spins, possibly spanning the entire
theoretically-allowed range given by the Kerr limit $|c\mathbf{S}/Gm^2| \leq
1$, where $\mathbf{S}$ is the spin angular momentum of the BH and $m$ is its mass.
These spin effects are apparent in the waveform templates, and using
non-spinning templates to search for spinning signals is sub-optimal, as we
quantify below.

Nonetheless, most previous compact-object binary searches with LIGO and
Virgo have used non-spinning templates to filter the
data~\cite{Abadie:2011kd, Aasi:2012rja, Colaboration:2011np,
  Aasi:2012rja}.  Although non-optimal, templates without spin effects
still have some overlap with spinning signals; searches with
non-spinning templates can still detect signals from spinning systems,
just at a lower rate. Furthermore, the only LIGO search that included
spin effects in the search templates~\cite{Abbott:2007ai} found that
the increase in the search background due to the increased number of
templates---required to cover the larger spin space---offsets the
gain in signal-to-noise ratio achieved by using them over non-spin
templates~\cite{VanDenBroeck:2009gd}. In order for spinning templates
to be effective in a search, further methods for distinguishing
between noise and signal, such as data-based consistency
tests~\cite{Allen:2004gu, Hanna:2008, Harry:2010fr}, would have to be
developed for spinning templates.

Here, we revisit the question of searching for BBHs using spinning templates.
Recent studies~\cite{Privitera:2013xza, Canton:2014ena} have demonstrated
methods for searching with (aligned) spinning templates that outperform the
best non-spinning template search in most regions of parameter space. In
Ref.~\cite{Privitera:2013xza}, they considered a search for BBHs in the mass
range $M_\mathrm{total} \in [10, 35]\Msun$ and mass ratio $1 \leq m_1/m_2 \leq
4$. The spin effects were modeled with an inspiral-merger-ringdown
phenomenological template family~\cite{Ajith:2009bn} that uses a single
effective-spin parameter
$\chi_\mathrm{eff} = {(m_1 \chi_1 + m_2 \chi_2)}/{(m_1+ m_2)}$
where
\begin{equation}
\chi_{1,2} = \frac{c\mathbf{S}_{1,2} \cdot \mathbf{\hat{L}}}{G m_{1,2}^2}
\end{equation}
are the dimensionless spin parameters of the BHs that are aligned
with the direction of the orbital angular momentum $\mathbf{\hat{L}}$. Due to
limitations in the regime of validity of the waveform model, the study in
Ref.~\cite{Privitera:2013xza} restricted the templates to span only
$\chi_\mathrm{eff} \in [-0.5, 0.85]$.  Analysing real initial LIGO detector
noise with simulated spinning signals added, the authors found that the
spinning template search improved the non-spinning one by 45\% for systems with
$M_\mathrm{total} \in [15,25]\Msun$ and $\chi_\mathrm{eff} \in [0.2, 0.85]$.
The study in Ref.~\cite{Canton:2014ena} considered spin effects in searches for
neutron-star--black-hole binaries, which we do not consider here.

We extend the work in Ref.~\cite{Privitera:2013xza} in several
significant ways.  Firstly, we describe inspiral-merger-ringdown
signals using effective-one-body (EOB) waveforms tuned to
numerical-relativity simulations~\cite{Taracchini:2013rva}. Those
waveforms describe double-spin BBHs and cover mass ratios $1 \leq
m_1/m_2 \leq 100$ and spins $\chi_{1,2} \in [-0.99,0.99]$.  Thus, we
can explore a larger BBH mass-parameter space, spanning
$M_\mathrm{total} \in [4, 50]\,\Msun$. Secondly, we demonstrate the
applicability of these methods to realistic aLIGO noise, filtering
from $f_\mathrm{low}=30$Hz, making the conclusions immediately
applicable to ongoing searches. We systematically and quantitatively
map out the regions of this extended parameter space in which
including spin effects in templates improve the search sensitivity. We
continue to consider only aligned spin templates here, as a search
using spin misaligned (precessing) templates is significantly more
challenging. We explore the question of precessing templates in a
companion work~\cite{Harry:2016ijz}.

In arriving at our results, we combine and improve upon two recent
implementations of the two template placement strategies mentioned
above; these implementations are described in
Refs.~\cite{Brown:2012qf, Harry:2013tca, Ajith:2012mn,
  Privitera:2013xza}. In Sec.~\ref{sec:bank_construction}, we review
these two template placement methods. Applying these methods along
with some additional computational enhancements, we demonstrate a
procedure for template bank placement that efficiently covers the
four-dimensional mass and (aligned) spin parameter space. In
Sec.~\ref{sec:sensitivity_comparisons}, we demonstrate the application
of this aligned-spin template bank in an end-to-end search pipeline on
simulated aLIGO noise, and quantify the gains of using the
aligned-spin bank in this pipeline relative to a template bank without
spin as function of the source parameters. In doing so, we address
directly the interplay between the offsetting effects of increased SNR
recovery and increased false alarm rates, both of which contribute to
the overall sensitivity of a search. We demonstrate the pipeline both
on Gaussian noise and initial LIGO noise recolored to the early aLIGO
spectrum. In Sec.~\ref{sec:conclusions} we consider the implications
of these results.

This study was completed prior to the start of the first observing run of
aLIGO (O1), which began on September 12, 2015 \cite{Abbott:2016blz}. Based on
the results of this study, the template-placement strategy described here was
used for creating the template bank used in the modeled search that found
GW150914 \cite{TheLIGOScientific:2016qqj}. Prior to the discovery of GW150914,
the existence of BBHs with component masses $\gtrsim 25\,\Msun$ were only
theorized from population synthesis models, as described above. As such, the
highest-priority target for stellar-mass BBH modeled searches was systems with
total mass $\lesssim 50\,\Msun$ leading up to the start of aLIGO
\cite{LVC:2015WhitePaper}. Higher-mass systems were to be searched for on a
longer time scale, as it was thought that more sophisticated search techniques
might be required to mitigate non-Gaussian transients that adversely affect
high mass templates. For this reason, in this study we consider BBHs with
total masses $M_{\mathrm{total}} \in [4,50]\,\Msun$ and component masses $\geq
2\,\Msun$. However, the methods described here can be extended to higher
masses. Indeed, in the engineering runs leading up to the start of aLIGO it
was found that standard search techniques could sufficiently mitigate
non-Gaussian transients for templates with total masses up to $100\,\Msun$.
The methods described here were therefore used to place a bank spanning
component masses $1$--$97\,\Msun$ and $M_{\mathrm{total}}<100\,\Msun$, with
dimensionless spins between $\pm0.985$ for component masses $>2\,\Msun$
\cite{TheLIGOScientific:2016qqj}.

\section{Template bank generation}
\label{sec:bank_construction}

In this section we describe a method to place an effectual template
bank of aligned-spin template waveforms to search for BBH
signals with component masses between 2 and 48 $M_\odot$, a maximum
total mass of 50 $M_\odot$, both component spins $\in [-0.99,0.99]$
and using the predicted 2015-16 advanced LIGO noise curve~\cite{Aasi:2013wya}. 
Our bank generation process relies on combining two existing algorithms, a geometric-based
aligned-spin algorithm, as described in Refs.~\cite{Brown:2012qf,Harry:2013tca}
and a ``stochastic'' algorithm, as described in Refs.~\cite{Ajith:2012mn,Privitera:2013xza}.
We begin by briefly reviewing the criteria that a template bank should fulfill to
be useful for gravitational-wave astronomy. We then describe the methods
used to place banks of non-spinning waveform filters in previous LIGO and Virgo searches and
demonstrate that these non-spinning banks are sub-optimal for our aligned-spin parameter space.
We then describe both the stochastic and geometric methods for placing banks of aligned-spin
waveforms and demonstrate the performance of template banks using each of these methods.
Finally, we introduce our new method of combining these approaches and demonstrate that this
combined approach generates a suitable, efficient template bank of aligned-spin BBH
template waveforms.

\subsection{Background}
\label{ssec:tmpltbanks_background}

Binary black hole mergers are described by 17 parameters; the component masses
($m_1$, $m_2$), the component spin vectors ($\mathbf{S}_{1}$,
$\mathbf{S}_{2}$), the eccentricity $e$ and phase of perihelion $\gamma$, the
right ascension and declination of the source ($\alpha$, $\delta$), the
distance $r$, the inclination angle $\iota$, the polarization phase $\psi$, the
orbital phase at coalescence $\varphi_c$ and the time at coalescence $t_c$.
Denote the set of these parameters $\allparams$. If $\allparams$ of a signal
$\signal$ are known \emph{a-priori}, then the likelihood ratio between the
signal-in-noise hypothesis $p(s|\signal)$ and the noise-alone hypothesis
$p(s|0)$ in a detector with stationary Gaussian noise is \cite{Finn:1992wt}:
\begin{equation}
 \label{eq:unnorm_likelihood}
 \mathcal{L}(\signal(\allparams)) = \frac{p(s | \signal(\allparams))}{p(s|0)} = 
  \exp\left[ \ip{\signal}{s} - \frac{\ip{\signal}{\signal}}{2} \right],
\end{equation}
where the inner product $\ip{\cdot}{\cdot}$ is defined as
\begin{equation}
\label{eqn:real_inner_product}
\ip{a}{b} \equiv 4 \Re \int^\infty_0 \frac{\tilde{a}^{*}(f) \tilde{b}(f)}{S_n(f)} df.
\end{equation}
We use $\tilde{a}$ to represent the Fourier transform of a time series $a$ and $S_n(f)$
denotes the one-sided noise power-spectral density of the data.

In reality, the parameters of astrophysical systems will not be known a-priori,
and searches must therefore be sensitive to signals at any location in the
17-dimensional parameter space.  Performing the matched-filter calculation at
every point in the full parameter space would be extremely computationally
prohibitive, and therefore a number of analytic approximations are used to
reduce the size of the parameter space.

The unknown distance $r$ to a signal simply scales the amplitude
of the waveform observed in the detector. We can therefore maximize over the distance
by writing the waveform as $\signal = A \template$, where $\template$ is the waveform at some
fiducial distance, and maximizing the likelihood over the amplitude $A$. Since the
likelihood ratio changes rapidly around the maximum, it is common practice to instead
maximize the log of the likelihood ratio. Doing so yields:
\begin{equation}
\label{eqn:maxloglikelihood}
\max_{r}\left[\log\mathcal{L}(\signal)\right] = \frac{\ip{\template}{s}^2}{\ip{\template}{\template}}.
\end{equation}

As in previous searches~\cite{Abbott:2009tt, Colaboration:2011np, Babak:2012zx},
we restrict ourselves to only considering non-precessing
binaries on circular orbits, and consider only the dominant spin-weighted spherical-harmonic mode,
the $(2,\pm2)$ mode. Making these assumptions will reduce detection efficiency to systems with precession,
eccentric orbits, or in which sub-dominant waveform harmonics are important. These effects have been investigated
in~\cite{Ajith:2012mn, Harry:2013tca, Huerta:2013qb, Capano:2013raa, Harry:2016ijz}, and may be
important for a small number of astrophysical systems. However, the first searches of Advanced LIGO
and Advanced Virgo data will make these assumptions and consider only dominant-mode, aligned-spin
non-eccentric waveform templates~\cite{Canton:2014ena}, and we restrict ourselves similarly here.

With these assumptions we have restricted to an 11-dimensional parameter space.
The remaining \emph{extrinsic} parameters --- inclination, polarization, sky location,
coalescence phase and coalescence time; collectively denoted $\extparams$---now enter the
gravitational waveform only as a constant time,
phase or amplitude shift~\cite{Babak:2012zx}. The constant amplitude shift cancels in the
maximized log likelihood of Eq.~\eqref{eqn:maxloglikelihood}. The constant phase shift $\phi$
and the coalescence time $t_c$ can be maximized over by re-defining the inner product such that
it is complex and has a time dependence~\cite{Allen:2005fk}:
\begin{equation}
 \label{eqn:inner_product}
 \ip{a}{b}(t_c - t_0) \equiv \int^\infty_0 \frac{\tilde{a}^{*}(f) \tilde{b}(f, t_0)}{S_n(f)} e^{-2\pi i f t_c} df.
\end{equation}
Here, $t_0$ is an arbitrary epoch. Taking the quadrature sum of the real and
imaginary parts of this complex inner product maximizes the likelihood ratio
over the unknown phase shift. We therefore define the matched-filter SNR $\rho$ as:
\begin{equation}
\label{eqn:snr_def}
\rho(t_c; \intparams|s) \equiv \frac{\left|\ip{\template(t_c; \intparams)}{s}\right|}{\sqrt{\ip{\template}{\template}}}.
\end{equation}
This can then be evaluated for all $t_c$ using a fast Fourier transform
routine~\cite{Allen:2005fk,FFTW} and numerically maximized over quickly.  The
remaining \emph{intrinsic} parameters, $\intparams = \{m_1,m_2, \chi_1,
\chi_2\}$, cannot be analytically maximized over. Instead, we create a set of
waveforms with varying values of these parameters --- a template bank --- and
filter all of the waveforms in the bank against the data. This template bank
should be constructed to have sensitivity over all of the parameter space of
interest.

When creating a template bank to use in searches for compact binary
coalescences we require some quantity that is a measure of the ``completeness''
of the bank.  This is used to judge if a bank adequately covers the parameter
space of interest.  For a single template with parameters $\intparams'$ this can
be expressed in terms of the \emph{overlap} between the template
$\template(\intparams')$, and a putative point with parameters $\intparams$ somewhere
in the parameter space of interest. The overlap is defined as:
\begin{equation}
\label{eqn:overlap}
\mathcal{O}(\intparams'|\intparams) \equiv \frac{\rho(\intparams'|\intparams)}{\rhoopt(\intparams)},
\end{equation}
where $\rho(\intparams'|\intparams)$ is the SNR with $s = \template(\intparams)$ and
\begin{equation}
\label{eqn:rho_opt}
\rhoopt(\intparams) \equiv \sqrt{\ip{\template(\intparams)}{\template(\intparams)}}
\end{equation}
is the \emph{optimal SNR}, which is the SNR when $\intparams' = \intparams$.
The overlap can take values $\in [0,1]$; we refer to $1 -
\mathcal{O}(\intparams'|\intparams)$ as the \emph{mismatch}.  The
\emph{effectualness} for a putative signal $h(\intparams)$ is defined as the
largest overlap between that signal and all templates in the template bank:
\begin{equation}
 \eff(\intparams) = \max_{\intparams'} \mathcal{O}(\intparams'|\intparams).
\end{equation}

There are a few possibilities for assessing the completeness of a template
bank. Traditionally the \emph{minimal match} criterion has been
used~\cite{Babak:2006ty}. The minimal match simply requires that a template
bank is constructed such that no putative signal anywhere in the parameter
space has an effectualness less than the minimal match.  When a bank fulfills
this minimal match criterion we refer to it as \emph{effectual}.  The minimal
match has traditionally been set to a value of 0.97 for previous LIGO and Virgo
searches~\cite{Babak:2012zx, Abbott:2009tt, Colaboration:2011np} and we follow
that approach here.  This number is chosen such that the signal loss due to the
discreteness of the template bank is not more than $\sim$10\%. This 10\% is
obtained by assuming every signal is recovered with a effectualness equal to
the minimal match and translating that into a loss of detection volume. In
reality, signals will be linearly distributed in
effectualness~\cite{Keppel:2013yia}, and in fact will tend to cluster towards
higher values of $\eff$ when the templates have some overlap, as is
inevitable in lattices in more than 1 dimension~\cite{Conway:1993}. Therefore
the loss in signal rate for an effectualness of 0.97 is smaller even than 5\%.
We note though that other errors, for example waveform modeling uncertainties
and data calibration uncertainty can also reduce the effectualness beyond the
minimal match criterion~\cite{Abadie:2010px}.

Before constructing template banks to cover the region of parameter space we are interested
in, we first define exactly how we will compute the completeness of the banks.
We will compute the effectualness between a large set
of signal points, drawn from all areas of the parameter space. This tests that the bank
is effectual for all mass and spin values being considered. We use $500\,000$ points drawn from a
distribution that is uniform in the spin magnitudes and log in the component masses. We use a log
distribution in mass because the mismatch between waveforms changes more rapidly at low
masses than at high masses.\footnote{As we do not expect real signals to be distributed
this way, we weight the signals to mimic a distribution uniform in component
mass when computing sensitive volume; see
Sec.~\ref{sec:sensitivity_comparisons} for more details.}

We plot the distribution of the effectualness
over the set of simulated signals in Fig.~\ref{fig:effectualness_cumhist}.
While the effectualness $\eff$ of each simulated signal is useful for identifying
areas of parameters space where a bank is not performing well, we wish to have
a single value that describes the performance of the entire bank. For that, we
do two additional comparisons.  First, we find the percentage of signals that
have $\eff < 0.97$, which is reported in Table
\ref{tbl:mass_spin_params}. Second, we compute a \emph{weighted mean
effectualness}. Previous studies have used an ``effective fitting factor'' to
assess the relative sensitivity of a bank, defined as \cite{Buonanno:2002fy}
$\left<\eff^3 \rhoopt^3\right>/\left<\rhoopt^3\right>$, where the mean is taken
over the set of simulated signals. The effective fitting factor gives an approximate
estimate of the fraction of signals that are detected by a bank assuming that
signals are distributed uniform in volume. However, we
have found that when considering a large range in masses, as we do in this
study, a few high-mass signals can dominate the effective fitting factor, even
when using a simulated signal that is log distributed in the component masses.
This is because the amplitude of a signal scales approximately by
$\mathcal{M}^{5/6}$, where $\mathcal{M} = M_\mathrm{total}(m_1 m_2 / M^2_\mathrm{total})^{3/5}$ is the \emph{chirp mass}
of the signal. Thus, the larger the chirp mass of a signal,
the larger its optimal SNR. To give equal weight to lower-mass signals, we
define the \emph{weighted mean effectualness} $\left<\eff_w\right>$
as:
\begin{equation}
\label{eqn:weighted_eff}
\left<\eff_w\right> = \frac{\left<(\eff\mathcal{M}^{-5/6}\rhoopt)^3\right>}{\left<(\mathcal{M}^{-5/6}\rhoopt)^3\right>}.
\end{equation}
These values are also reported in Table \ref{tbl:mass_spin_params}.

We use waveform models from double-spin BBH mergers built within the 
effective-one-body formalism, notably the non-precessing (or ``aligned'') 
spin templates developed in Ref.~\cite{Taracchini:2013rva}. The waveforms 
describe the full coalescence process, i.e., inspiral, merger and ringdown, 
but include only the main spin-weighted spherical-harmonic mode, i.e., 
the $(2,2)$ mode. Henceforth, we shall refer to those waveforms as 
SEOBNRv2 templates. 

\subsection{Non-spinning template banks}
\label{ssec:nonspin_banks}

\begin{table}
  \begin{tabular}{c|c|c|c}
    Template bank & Size & \parbox[t]{2cm}{$\%$ of signals \\with $\eff < 0.97$} & $\left<\eff_w\right>$ \\
    \hline
    \hline
    Non-spinning geometric & 7\,734 & 50 & 0.738\\
    Non-spinning combined & 8\,935 & 51 & 0.737\\
    Aligned-spin geometric & 57\,177 & 8.5 & 0.954\\
    Aligned-spin stochastic & 64\,318 & 0.01 & 0.970\\
    Aligned-spin combined & 60\,766 & 0.01 & 0.970\\
  \end{tabular}
  \caption{The sizes of the various template banks constructed in Sec.~\ref{sec:bank_construction}.
  The non-spinning and aligned-spin combined banks refer to banks constructed using both the geometric
  and stochastic algorithms, as discussed in Sec.~\ref{ssec:combined_algrthm}.
  }
  \label{tbl:mass_spin_params}
\end{table}

The template placement algorithms that were used for creating banks of non-spinning signals for previous
compact-object binary searches in LIGO and Virgo data are
described in 
Refs.~\cite{Sathyaprakash:1991mt, Poisson:1995ef, Balasubramanian:1995bm, Owen:1995tm, Owen:1998dk, Babak:2006ty, Cokelaer:2007kx}.
This method requires an analytical prediction of the mismatch
between two nearby waveforms to create a parameter-space metric describing how far apart
two points must be in the parameter space of the two masses before their overlap drops to
a specified value. This approximation is only valid for overlaps close to 1, but has been
found to be a very useful tool when creating banks with a minimal-match of 0.97.
For bank construction the parameter space metric must be constant---or
almost constant---over the whole parameter space. Currently such a metric only exists
for the inspiral-only ``TaylorF2'' waveform approximant and requires the assumption that
the termination frequency of the waveform will be constant over the parameter
space~\cite{Babak:2006ty,Brown:2012qf,Keppel:2013kia}.
With this TaylorF2 parameter space metric it is possible to place a regular hexagonal lattice in the
two-dimensional, non-spinning parameter space that covers the entire space at a desired minimal
match~\cite{Cokelaer:2007kx}.

Using this traditional non-spinning template bank construction algorithm with the TaylorF2
parameter space metric, we generate a template
bank of waveforms within the mass range described above and neglecting spin effects.
Then, modelling all of the templates with the SEOBNRv2 waveform model, we
compute the effectualness of this non-spinning template bank to SEOBNRv2 aligned-spin signals. 
The result of this is plotted in Figs.~\ref{fig:effectualness_scatter} 
and \ref{fig:effectualness_cumhist}. The number of templates in this
bank---7\,734---and all other banks described in this section are listed in Tab.~\ref{tbl:mass_spin_params}.
It can clearly be seen from Fig.~\ref{fig:effectualness_cumhist} that a large number of signals
were recovered with the non-spinning bank with effectualness less than 0.97.
Indeed, roughly 30\% of signals have $\eff < 0.9$. We translate this into detection
volume, compare this with our aligned-spin banks and assess performance in
different regions of the mass-spin parameter space later in the work. However, this plot
indicates that an aligned-spin bank could offer a significant improvement in detection rate.

\subsection{Geometric algorithm}
\label{ssec:geom_algrthm}

In Refs.~\cite{Brown:2012qf,Harry:2013tca} the authors extended the non-spinning geometric approach to
include aligned-spin signals for binary neutron star and neutron-star black-hole template bank placement,
allowing for the higher dimensionality of the aligned-spin
parameter space. However, geometric placement
requires a parameter space that is globally flat. Efficient lattice algorithms are not
known for general, intrinsically curved, parameter spaces~\cite{Conway:1993}.
To obtain a parameter space metric that is constant over the whole parameter space, we must use the
inspiral-only TaylorF2 parameter space metric, and assume that all waveforms terminate at the same
fixed frequency~\cite{Brown:2012qf}. For binary neutron-star signals
this is a valid approach as the merger occurs at a frequency outside
of the range of sensitivity for ground-based interferometers~\cite{Brown:2012qf}. However, for BBH signals, the
parameter space is large and these assumptions are not valid for template bank placement,
as we will demonstrate. For non-spinning
bank placement this approach is equally invalid, but generally it produces template over-coverage in the
high-mass parameter space. As the template density is low anyway in the high-mass parameter space, this
results in effectual template banks with only a small amount more template waveforms than are needed.
However, when considering aligned-spin systems there is a strong degeneracy between mass ratio and
the spins, which can be broken when waveforms terminate at different frequencies. As the geometric
approach cannot take this into account it can create template banks that are not effectual, as we
will show below.

\begin{figure*}
  \begin{center}
  \begin{minipage}{0.27\paperwidth}
  \begin{subfigure}{\textwidth}
    \includegraphics[width=\textwidth]{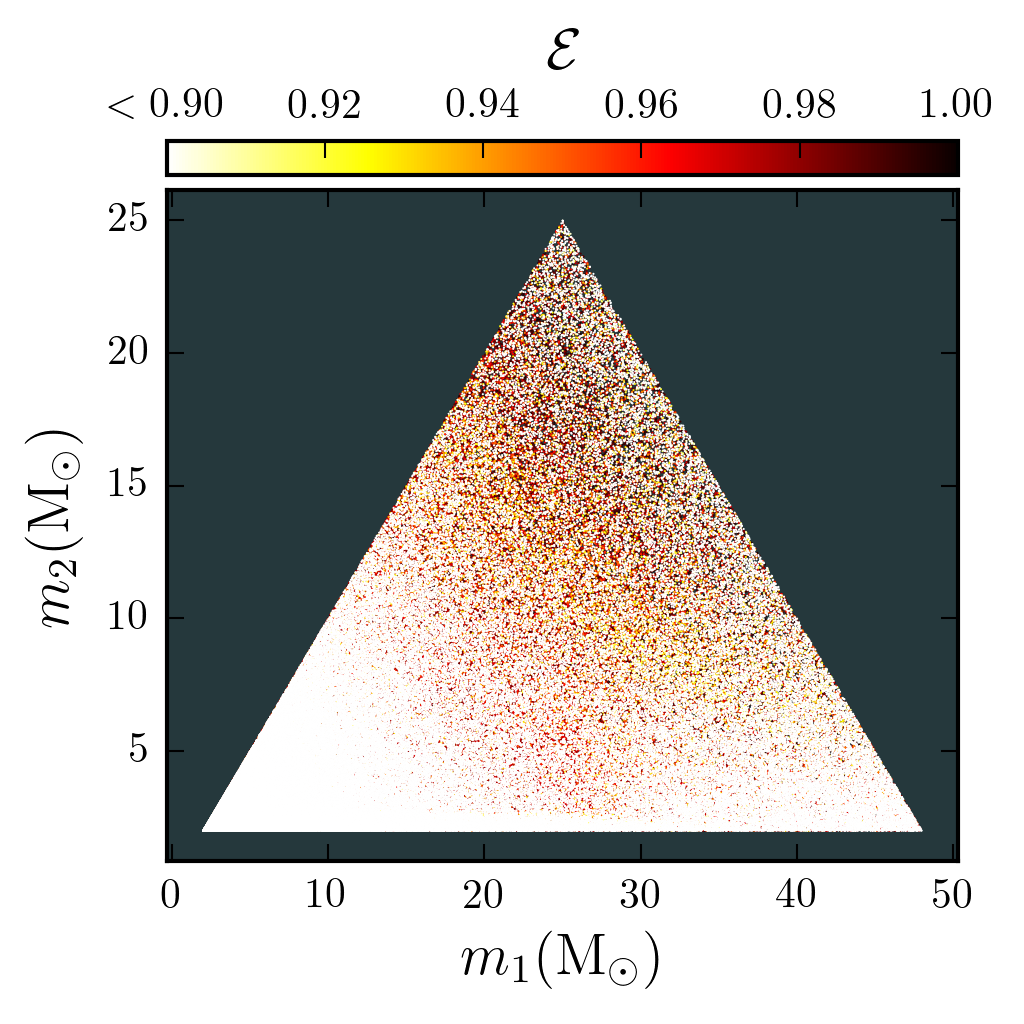}
    \\
    \includegraphics[width=\textwidth]{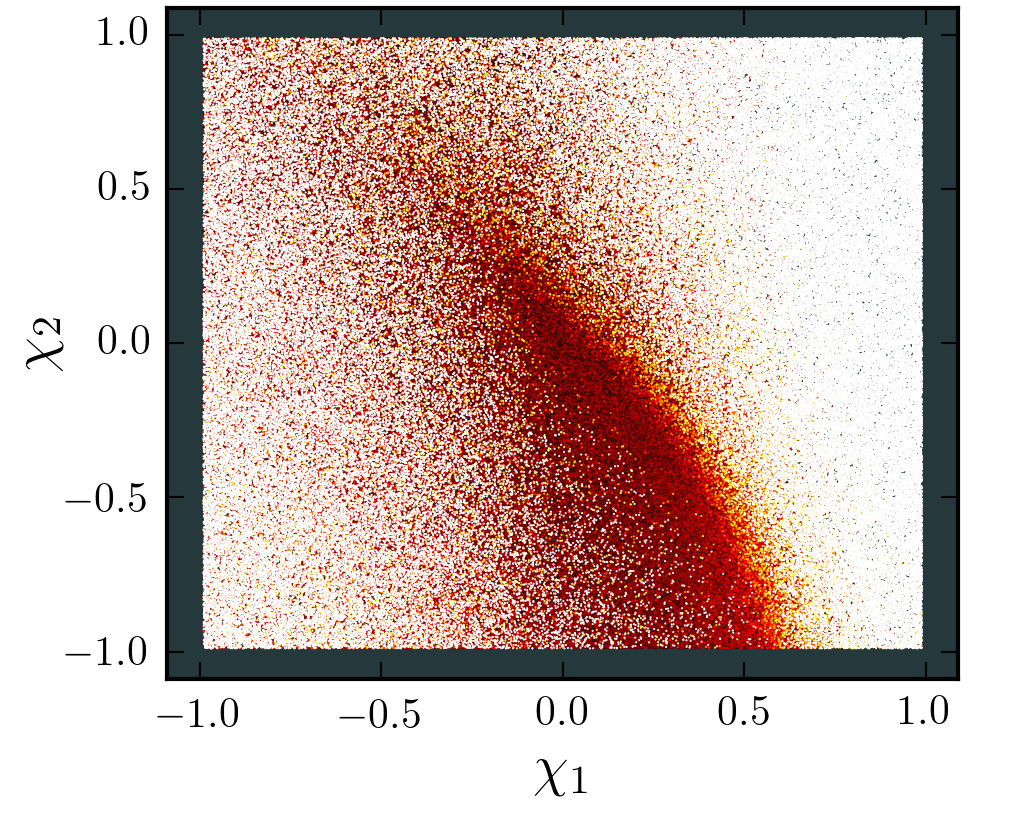}
    \caption{Non-spinning: geometric-only}
  \end{subfigure}
  \end{minipage}
  \begin{minipage}{0.27\paperwidth}
  \begin{subfigure}{\textwidth}
    \includegraphics[width=\textwidth]{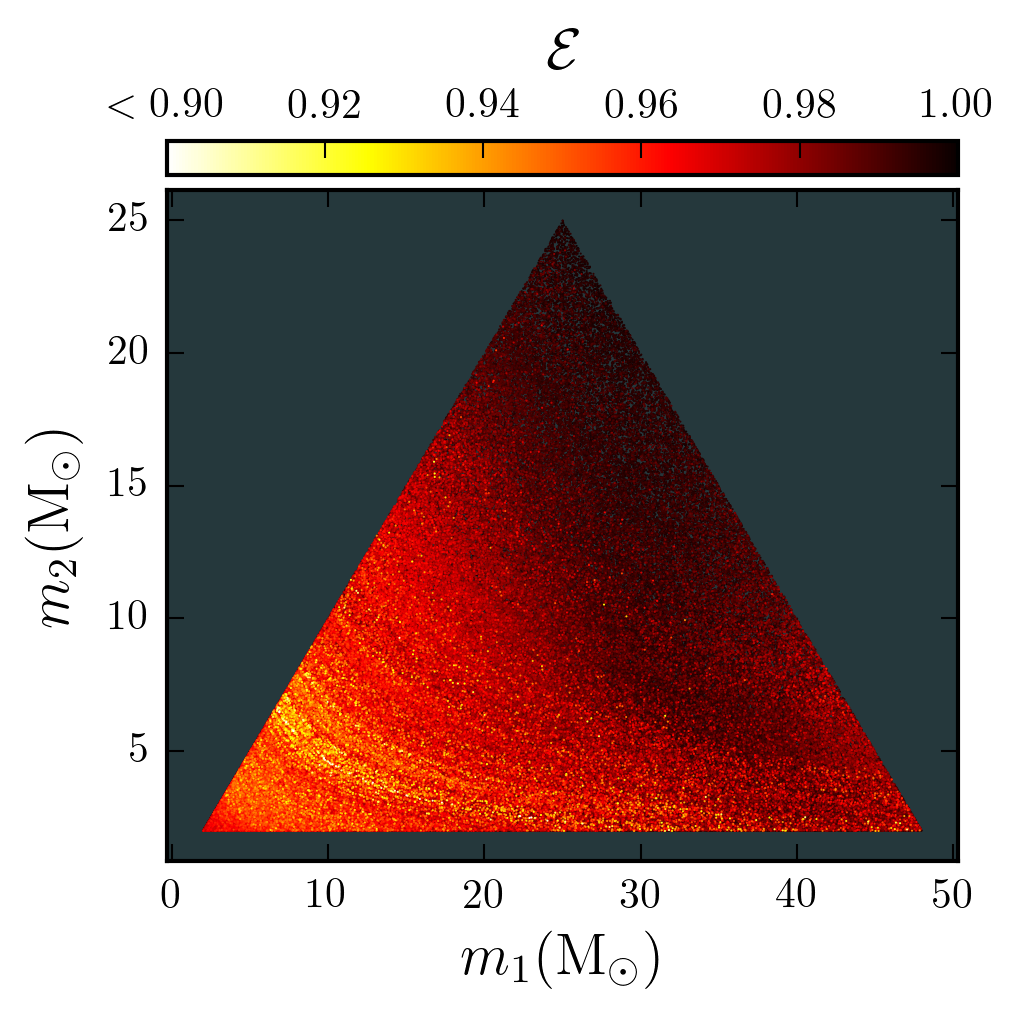}
    \\
    \includegraphics[width=\textwidth]{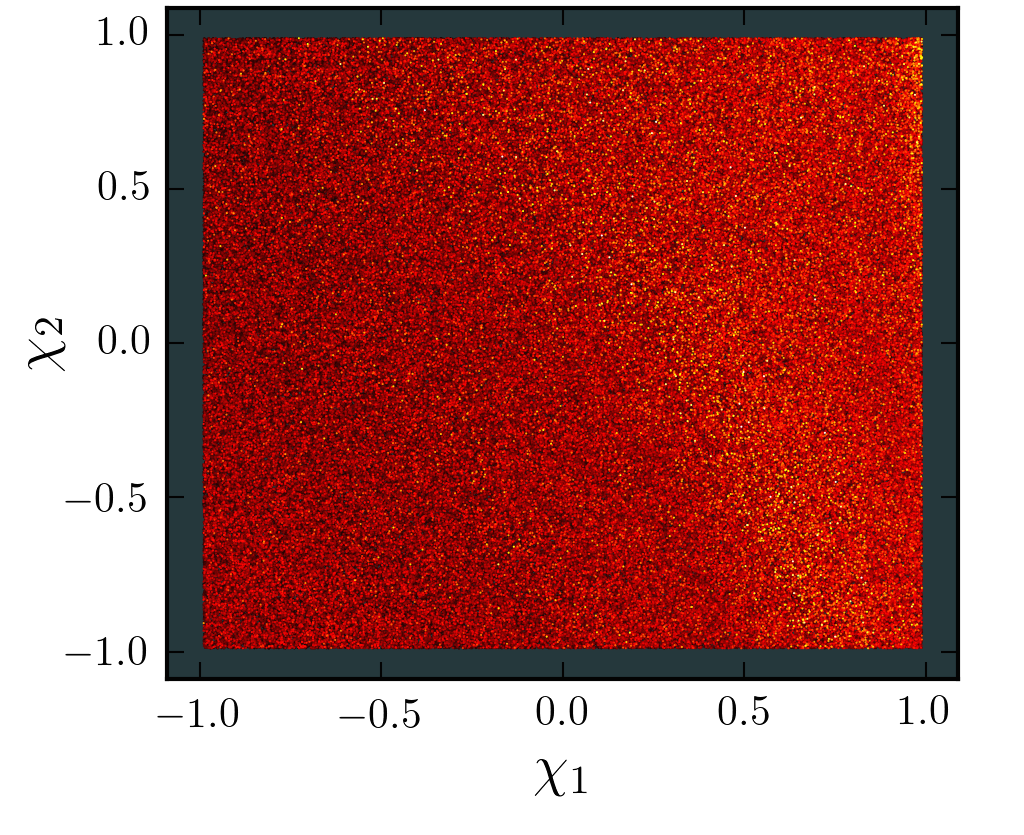}
    \caption{Aligned-spin: geometric-only}
  \end{subfigure}
  \end{minipage}
  \begin{minipage}{0.27\paperwidth}
  \begin{subfigure}{\textwidth}
    \includegraphics[width=\textwidth]{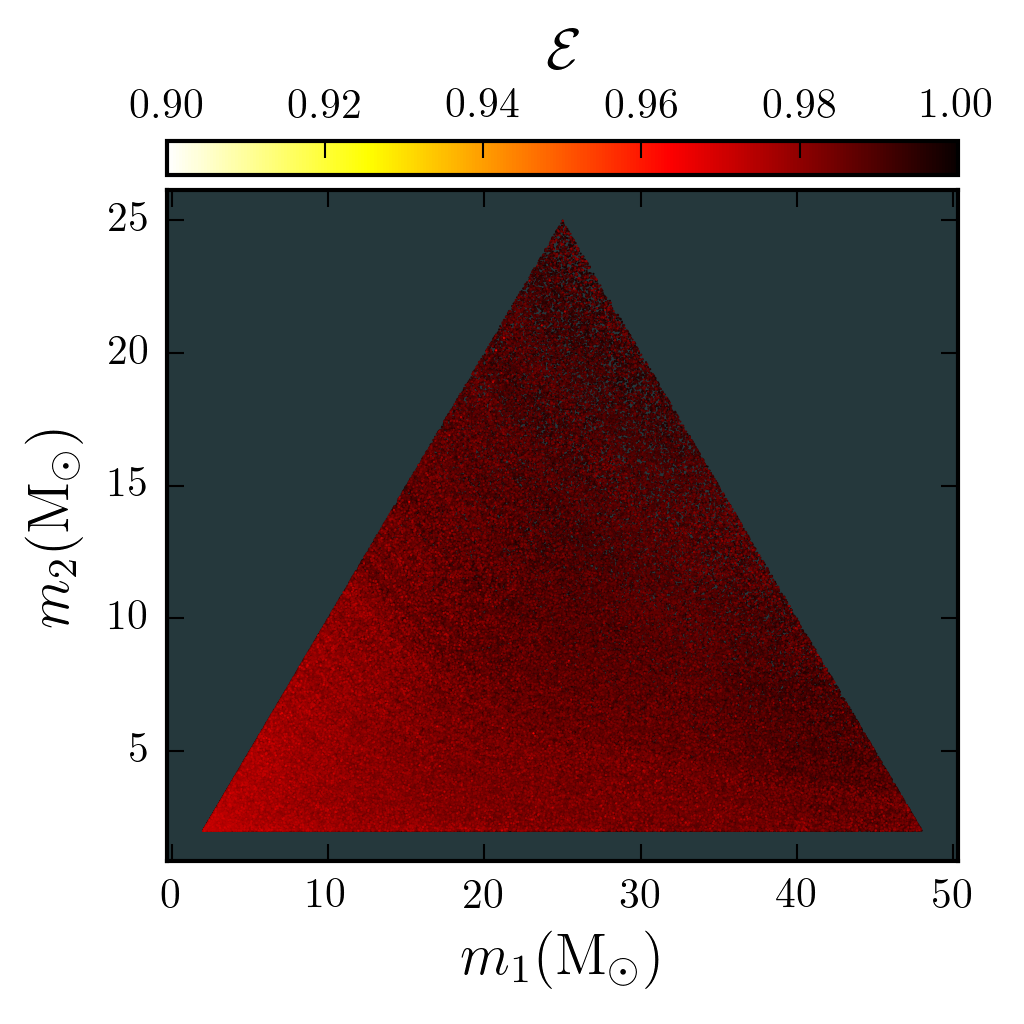}
    \\
    \includegraphics[width=\textwidth]{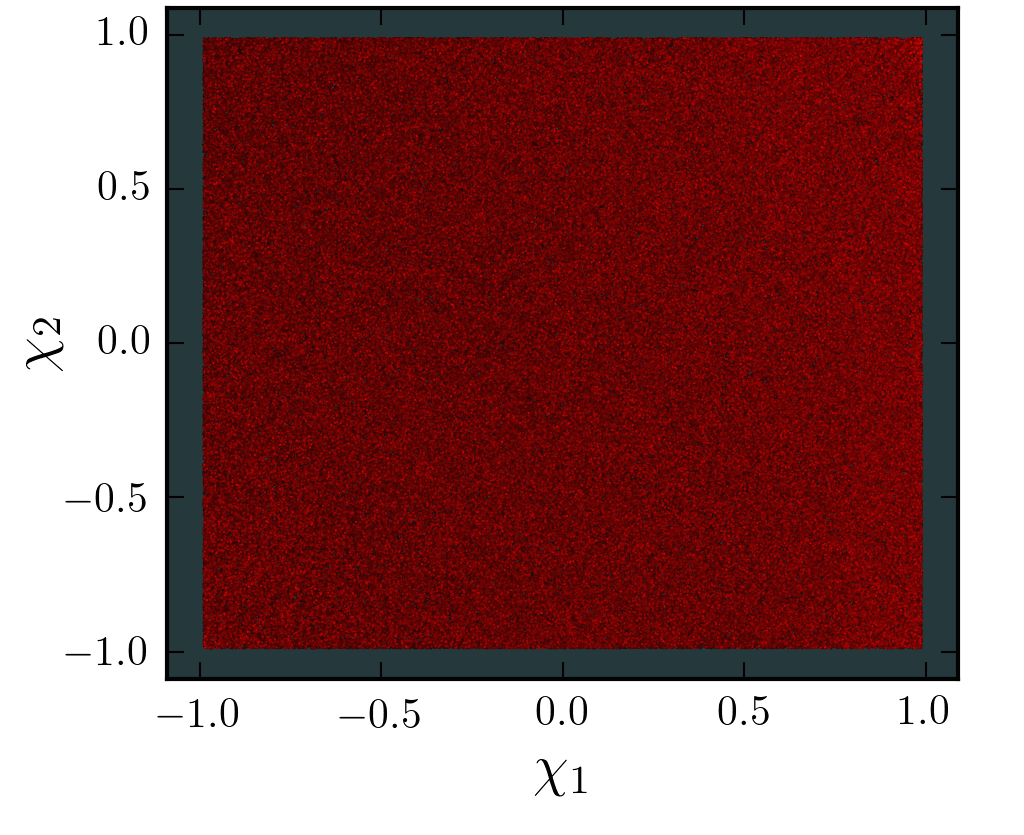}
    \caption{Aligned-spin: combined}
  \end{subfigure}
  \end{minipage}
  \end{center}
  \caption{
    Effectualness ($\eff$) as a function of $m_1, m_2$ (top) and $\chi_1,
    \chi_2$ (bottom) of the non-spinning bank, the aligned-spin geometric-only
    bank, and the aligned-spin with geometric and stochastic placement bank.
    Each point represents a simulated signal.
    }
  \label{fig:effectualness_scatter}
\end{figure*}

\begin{figure}
  \centering
  \includegraphics[width=\columnwidth]{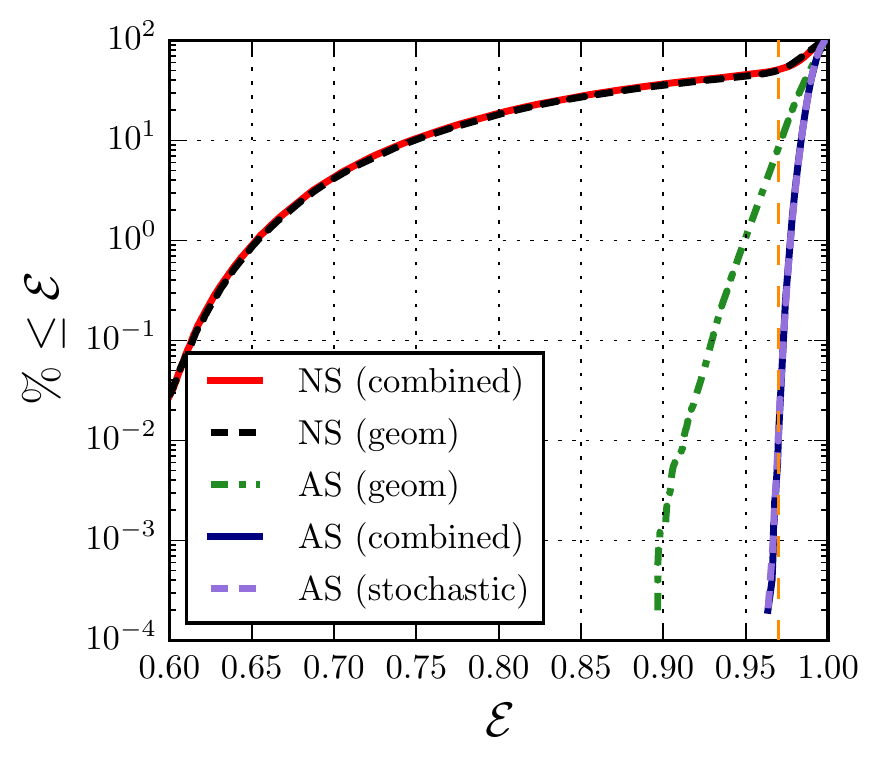}
\caption{
    Cumulative histogram of the effectualness ($\eff$) of each bank. The y-axis shows the
    percentage of simulated signals that have $\eff \leq$ the value given on the x-axis.
    The vertical orange line shows the desired minimal-match of the banks ($= 0.97$).
    }
\label{fig:effectualness_cumhist}
\end{figure}

We construct a geometric aligned-spin bank to cover the parameter space using the TaylorF2
aligned-spin metric and choosing a fixed value for the waveform cut-off frequency of 1100Hz.
We then test effectualness using the same set of points as for the non-spinning bank and again
modelling the aligned spin templates and signal waveforms using the SEOBNRv2 waveform model.
The results of this
are also shown in Figs.~\ref{fig:effectualness_scatter} and \ref{fig:effectualness_cumhist}.
This bank contains 57\,177
templates. We can see that the effectualness for this aligned-spin bank
is much closer to the desired minimal-match criterion of $> 0.97$, however there are
regions of parameter space where the minimal match can be as low as 0.9. The geometric approach 
offers us an efficient way of covering the low-mass parameter space, but
is not effectual everywhere when considering broad parameter spaces.

\subsection{Stochastic algorithm}
\label{ssec:stoch_algrthm}

An alternative method for placing banks of aligned-spin systems is the stochastic algorithm. In
this method one randomly places a large set of points in the parameter space and then iterates
over these points accepting each point into the template bank only if its overlap with all
points already accepted to the template bank is less than the minimal match. This method was
first proposed in the context of the LISA space-based
detector~\cite{Harry:2009ea, Babak:2008rb}, and
has been adapted to the problem of aligned-spin template placement for LIGO and Virgo searches
in Refs.~\cite{Ajith:2012mn,Privitera:2013xza}.
This method can only guarantee that all points of the parameter space are covered to the minimal-match 
criterion if an infinite number of seed points are used. Therefore some approximation to
the minimal-match criterion must be used, such as limiting the total number of seed points, or terminating the
iterative process after a specific number of points have been rejected in succession. The stochastic
algorithm uses more templates to cover a parameter space than the geometric approach, and can
be computationally expensive when the overlaps are computed explicitly.
However, this method offers a general approach that can be used to place
a template bank for any parameter space.

Applying the stochastic method directly to our aligned-spin parameter space is computationally
expensive. In order to optimize this process and speed up the generation of an aligned-spin
template bank in this parameter space we make use of two new optimizations, in addition to
the methods described in Refs.~\cite{Ajith:2012mn,Privitera:2013xza}.

The value of the frequency spacing used in the matched-filter integral ($df$)
is normally chosen to be $1/L$, where $L$ is the closest power-of-2 that is greater than the length
of the waveform (in seconds). This is sufficient to measure the overlap between two waveforms
in a time window of $L$ seconds. However, for bank generation we are only interested in the \emph{maximum}
overlap between two waveforms. If the two waveforms are aligned so that their peak amplitudes occur at the same
time, the maximum overlap is near to the time point corresponding to 0 displacement
between the two waveforms. Therefore, we can increase the value of $df$, which reduces the
cost of the inverse Fourier transform used to obtain the overlap as a function of time. To be
sure that the value of $df$ is not set so large that an incorrect overlap is obtained we compute
the overlap using some initial value of the frequency spacing ($df_0$) and also compute a
second overlap using $df_0/2$.
If both overlaps agree to within 1\%,
or if either overlap is less than four times the difference between the minimal match and
unity---0.88 in this case---we use that value. Otherwise the overlap is computed again
at $df_0/4$ and compared to the value obtained at $df_0/2$.
This process continues iteratively until the value of
the overlap converges. In our testing we found that $df_0 = 4$Hz was a suitable choice and
that is used in the numbers and results quoted below. In this manner, we reduce the cost of
computing overlaps, and can quickly assess cases where the overlap between two waveforms
is small. This significantly reduces the computational cost of the
stochastic bank. A similar approach has also been employed in the context of parameter
estimation to achieve a similar speedup in computational time~\cite{Purrer:2015nkh}.

In the approach described in Refs.~\cite{Ajith:2012mn,Privitera:2013xza} the stochastic
step must be parallelized due to computational cost. This is done by
splitting the parameter space into a number of non-overlapping chirp-mass bins and running the stochastic bank generator individually on each chirp-mass region.
Each individual job places points until a specified number of points, 100\,000 in
our case, were rejected while accepting the last 10 templates
into the bank. This parallelization results in some over coverage along the
chirp-mass boundaries and so the number of chirp-mass bins must be chosen to balance
this over coverage against the computational cost of generating the bank. We investigated
using varying numbers of non-overlapping chirp-mass regions and found that in this case 25
regions provided the
best balance between these two factors.
However, we note that the majority of the computational cost associated with the
stochastic algorithm is spent accepting the final
small number of templates~\cite{Harry:2009ea}.
We therefore organize the stochastic placement in
two steps. First, we run a single instance of the stochastic generator, covering the full
parameter space, but have it terminate when only 2500 points
have been rejected in accepting the last 10 templates. Then this semi-complete stochastic bank
can be used as a seed to the parallel generation to ensure completeness while minimizing double
coverage along the boundaries of the chirp-mass bins.

We construct an aligned-spin stochastic template bank using these methods and again test
the effectualness of this bank using the same set of test points as before.
This bank contains 64\,318 templates, which can be compared with the 57\,177 templates that
the geometric algorithm placed. However, in Fig.~\ref{fig:effectualness_cumhist} we
can see that the stochastic bank more completely covers the parameter space, with only 0.01\%
of points in the parameter space having an effectualness less than 0.97.

\subsection{A combined geometric-stochastic approach}
\label{ssec:combined_algrthm}

The geometric and stochastic placement methods both offer their own advantages and disadvantages when
applied to bank placement for BBH signals for aLIGO and advanced Virgo. Previous works have always
focused on using one method or the other~\cite{Harry:2013tca,Privitera:2013xza}.
Here we propose that the
best results are found if we combine the two methods together. Specifically we propose that
template banks for BBH searches in aLIGO and advanced Virgo take the following approach. First,
generate a bank using the aligned-spin geometrical lattice up to some total mass for which the
placement is valid. For the early aLIGO noise curve that we consider here, we have empirically
determined that a boundary on the total mass of 6 $M_\odot$ provides a suitable boundary at which
to stop the geometric approach. Second, we use the stochastic algorithm, except, instead of starting
with an empty template bank, we start with the aligned-spin geometric bank and test points in parameter
space against this ``seed'' bank. No mass limits are given to the stochastic algorithm and it ensures 
that the full parameter space is covered, including any ``holes'' that might have been left
in the original geometric template bank.

As with the other banks in this section we compute effectualness using the set of test points
described at the top of this section. This bank contains 60\,766 points and
the distribution of effectualness can be seen in Figs.~\ref{fig:effectualness_scatter} and
\ref{fig:effectualness_cumhist}. We can see that in
this case only 0.01\% of the test points have a $\eff < 0.97$; the lowest value
in our set of 500\,000 points is at a minimal match of 0.964. As mentioned
above this is consistent with the stochastic algorithm, which cannot guarantee that 100\% of
points has minimal match greater than some threshold.
As a balance between template number and signal recovery we recommend that this combined method
be used for producing banks of aligned-spin BBH template waveforms.

Finally, for completeness, we also generate a \emph{non-spinning} bank combining
both geometric and stochastic placement as described above. This allows us
to make direct comparisons between the non-spinning and aligned-spin banks,
generated using the same combination of the geometric and stochastic algorithms
in the remainder of the paper. This bank contains 8\,935 templates and is also plotted
in Fig.~\ref{fig:effectualness_cumhist}. The performance of this bank when searching
for aligned-spin signals is largely
indistinguishable from the non-spinning bank generated using the traditional, geometric,
method.

\section{Relative gain over non-spinning bank}
\label{sec:sensitivity_comparisons}

Having arrived at an effectual aligned-spin template bank, we can now
investigate the gain in sensitivity by using this bank in place of a
non-spinning bank. To do this we estimate the \emph{sensitive volume}, $\vol$, for a search using
each bank. The sensitive volume can be thought of as the mass-averaged volume in which a signal
would be detected above a specific value of some ranking statistic $\rstat$. This can be written
explicitly as
\begin{equation}
\label{eqn:sensitive_v}
\vol(\rstat) = \int \epsilon(\rstat; \intparams, \extparams, \spatialV) q(\intparams, \extparams, \spatialV)
            \mathrm{d}\spatialV \mathrm{d}\intparams \mathrm{d}\extparams.
\end{equation}
Here, $\spatialV$ is the three-volume of space. For compactness, we let
$\allparams = (\intparams, \extparams)$. The function
$q(\allparams, \spatialV)$ is the distribution of signals in the universe;
i.e., it is the number of signals that exist in the universe per unit time.
The efficiency $\epsilon(\rstat; \allparams, \spatialV)$ is the fraction of
those signals that can be detected by the search at the given $\rstat$.
For comparing pipelines, we will find it convenient to normalize $q$ such that
if we integrate it over some fiducial spatial volume $V_{\max}$, we have:
\begin{equation*}
\iint \int\limits_{V_{\max}} q(\allparams, \spatialV) \mathrm{d}\spatialV \mathrm{d}\allparams = V_{\max}.
\end{equation*}
We choose $V_{\max}$ such that the efficiency of the search is $0$ to any
signal outside of $V_{\max}$.

The sensitive volume is dependent on the distribution of signals in the
universe, $q$. For BBHs with total masses $\leq 50\,\Msun$, the detectors are
sensitive out to a maximum distance of $\sim1\,\mathrm{Gpc}$ assuming the
early advanced LIGO PSD. Over these distances we can assume an isotropic
distribution of signals \cite{Abadie:2010cf}; i.e., we assume that signals are
distributed uniformly in inclination, sky-location and orientation. The
detection of GW150914 confirmed that BBHs with $M_{\mathrm{total}} \approx
65\,\Msun$ exist \cite{TheLIGOScientific:2016wfe}, but beyond that, the mass
and spin distribution of BBHs is largely unknown. For simplicity, in this
study we assume an astrophysical prior that is uniform in component mass
between $2$ and $25\,\Msun$ and uniform in aligned-spin magnitude. However, in
the following sections we also produce sensitive volumes as a function
component mass and spin, to mitigate the effects of choosing an inaccurate
prior.

With these assumptions of the distribution of signals, we show in the appendix
that the sensitive volume is approximately [see Eq.~\eqref{eqn:logweightedMCV}]:
\begin{equation}
\vol(\rstat) \approx 4 \pi\sum_{i}^{N}  \tilde{w}_i \left[r_{\min,i}^3 + 3\Theta(\rstat_i) \Delta r_i r_i^2\right],
\end{equation}
where the sum is over the same simulated signals we used in Sec.
\ref{sec:bank_construction} to assess the effectualness of each bank. The
function $\Theta(\rstat)$ equals one if a simulated signal has a ranking
statistic $\geq \rstat$ and zero otherwise; the normalized weights
$\tilde{w}_i$ are needed to convert from the simulated signals' mass
distribution ($\log$ in the component masses) to our assumed astrophysical
distribution (uniform in the component masses) [see
Eq.~\eqref{eqn:normed_weights}].  Each signal's distance $r_i$ is drawn
uniformly between distance bounds $r_{\max,i} - r_{\min,i} \equiv \Delta
r_i$.\footnote{These bounds are different for each signal; see the Appendix for
details.}

The ranking statistic $\rstat$ is used to determine the likelihood
that candidate events are GWs. We choose a threshold $\threshrstat$ such that the
probability of mis-identifying noise as a detection---the \emph{false alarm
probability}---is small. For larger template banks, searching larger regions of
parameter space, the rate of background triggers above a given value of
detection threshold increases. Therefore a template bank covering the
aligned-spin parameter space has a larger false alarm probability $\fap$ at
fixed value of detection threshold than a non-spinning template bank. To keep
the false alarm probability fixed, the threshold at which a detection could be
claimed must therefore increase for the aligned-spin bank.

If the detector data were stationary Gaussian noise, the optimal (in the Neyman-Pearson sense) ranking
statistic would be SNR. Real gravitational-wave
detector data is not Gaussian.  Due to the presence of
non-Gaussian transients (\emph{glitches}), signal-based vetoes are needed to
separate glitches from real signal candidates~\cite{Aasi:2012wd,Aasi:2014mqd}.
Several signal-based vetoes have
been proposed \cite{Allen:2004gu,Hanna:2008,Harry:2010fr}.
The signal-based veto used in the most recent searches for compact-object binary mergers
\cite{Colaboration:2011np, Aasi:2012rja}, and the one we adopt here, is
the $\chi^2$ test first proposed in Ref.~\cite{Allen:2004gu}.
This statistic splits the template $\template$ into $p$ non-overlapping frequency bins such
that each sub-template $\template_i$ contributes an equal amount to the
SNR if the data exactly matches the template. Filtering each $h_i$ with the
data, we can test how well the frequency evolution of the SNR matches the
expected evolution with the statistic:
\begin{equation}
\label{eqn:defChisq}
\chi^2 = \frac{p}{\ip{h}{h}} \sum_{i=1}^{p} \left|\ip{h_i}{s} - \frac{\ip{h}{s}}{p}\right|^2.
\end{equation}
In Gaussian noise, this statistic is $\chi^2$ distributed with $2p-2$ degrees
of freedom. The greater the mismatch between the data and the template
the larger the $\chi^2$. This $\chi^2$ statistic as well as the SNR is then
used to compute a detection statistic \cite{Colaboration:2011np}
\begin{equation}
\label{eqn:newsnr}
\rhonew = \left\{
\begin{array}{cl}
\rho & \text{for $\chi_r^2 \leq 1$}, \\
\rho
\left[\frac{1}{2} \left(1 +
\left(\chi_r^2\right)^{\!3}\right)\right]^{-1/6}
& \text{for $\chi_r^2 > 1$,}
\end{array} 
\right.
\end{equation}
where $\chi^2_r = \chi^2/(2p-2)$ is the \emph{reduced} $\chi^2$. In this study
we use $p=16$, as used in the search for compact-object binary mergers with
total mass less than 25 $M_\odot$ in initial LIGO and initial Virgo's last
observing runs \cite{Colaboration:2011np}. We note that another choice for the
number of $\chi^2$ bins can produce better sensitivity. In the search over the
initial LIGO and Virgo data for binaries with total mass greater than
$25\,\Msun$ $p=10$ was used \cite{Aasi:2012rja}. However, further tuning of
this parameter is outside the scope of this paper.

It has been shown \cite{Babak:2012zx} that re-weighting the SNR via Eq.
\eqref{eqn:newsnr} down-weights glitches sufficiently such that the
distribution of $\rhonew$ in noise is close to that of SNR in Gaussian noise.
Conversely, $\rhonew \approx \rho$ for signals, as long as the mismatch between
signals and templates is small. Therefore, the re-weighted SNR allows searches
for compact-object mergers to reach comparable sensitivities to the ideal case
where the detectors' data is Gaussian. However, if templates do not match
signals well, as in the case of the non-spinning bank searching for spinning
signals, then the $\rhonew$ of those signals will be less than the raw SNR
$\rho$.  This reduces the sensitive volume of the non-spinning bank compared to
what would be obtained if only $\rho$ was considered, as has been done in many
other studies~\cite{Brown:2012qf,Privitera:2013xza,Harry:2013tca}.

Searches for BBHs also require that candidate gravitational wave triggers occur
in multiple detectors with the same mass and spins within the light-travel time
between the detectors. In that case, the \emph{network} re-weighted SNR for the
search is computed from the quadrature sum of the single-detector $\rhonew$.
The network re-weighted SNR is the ranking statistic $\rstat$ we use to compute
the sensitive volume.

In the following section we compare the sensitive volumes of the aligned-spin
bank to the non-spinning banks. In Sec.  \ref{ssec:gaussian_comparison} we use
simulated Gaussian noise for each detector. In Sec.
\ref{ssec:recolored_comparison} we run the full modern search pipeline
described in Ref.~\cite{Canton:2014ena} on a subset of the simulated signals
using data from initial LIGO's sixth science run (S6) recolored to resemble the
predicted sensitivity of aLIGO's first observing run.

\subsection{Gaussian noise}
\label{ssec:gaussian_comparison}

As we will show below, the sensitive volume is strongly dependent on our choice
of astrophysical prior. Higher-mass systems tend to dominate the sensitive
volume estimate because they emit more power in gravitational waves. This is
particularly true when assuming a prior that is uniform in component mass, as
we have done here. We will obtain misleading results if our prior is wrong, which
is likely given the large uncertainty in the mass and spin distribution of BBHs.

To mitigate the effect of our choice of prior, we wish to explore how the
sensitive volume changes across masses and spins. Doing so requires a large
number of simulated signals, as the variance in the volume estimate increases
the more we sub-divide the parameter space. However, adding more than a few
thousand simulated signals to real detector data and analyzing with the full
search pipeline is computationally expensive, as it requires finding the
overlap between every template and every simulated signal to find the best
matching template in a particular realization of noise.

Instead, in this section we use Gaussian noise to approximate the average
sensitive volume. By definition, the most effectual template to a signal will
be the template that has the largest SNR when averaged over several
realizations of Gaussian noise. We therefore do the following: we only filter
each simulated signal with its most effectual template in $16$ realizations of
simulated Gaussian noise in each LIGO detector.  We find the network
re-weighted SNR $\rstat$ in each realization, then average over the
realizations to get a measurement of the expectation value of $\rstat$,
$\left<\rstat\right>$. We use this to compute the sensitive volume. This allows
us to use all $500\,000$ simulated signals from the previous section for
computing sensitive volume.

Another advantage of using Gaussian noise is we can analytically estimate the
increase in false alarm probability at fixed $\rstat$ of the aligned-spin bank
as compared to the non-spinning bank. In real data the distribution of
re-weighted SNR is not the same for all templates. The shorter the bandwidth of
a template in the frequency domain, the larger its overlap with non-Gaussian
transients. This causes shorter-bandwidth templates to produce triggers with
large values of $\rstat$ at a higher rate than larger-bandwidth templates
\cite{Babak:2012zx}.  To account for this, searches have binned results by
various parameters when estimating false alarm rate, then combined results over
the bins \cite{Colaboration:2011np, Aasi:2012rja}. The choice of parameter to
use and the size of bins adds an additional complication when comparing
sensitivity, and is dependent on the noise.

However, in Gaussian noise, the SNR of every template is $\chi$ distributed
with two degrees of freedom. We therefore do not need to worry about binning
results by parameters. Furthermore, in Gaussian noise we can analytically
estimate the increase in false alarm probability due to the larger parameter
space covered by the aligned-spin bank. Let us assume that every template is
independent of each other. With this assumption, if we have $N_t$ templates in
a bank, the probability of getting one or more single-detector triggers
with an SNR $\geq \rho$ is:
\begin{equation}
P(\rho | N_t) = 1 - C_{\chi}(\rho)^{N_t},
\end{equation}
where $C_{\chi}(\rho) = 1 - e^{-\rho^2/2}$ is the cumulative distribution
function of the $\chi$ distribution with two degrees of freedom. In Gaussian
noise with no signals, $\rhonew \approx \rho$; with two detectors, $\rstat
\approx \sqrt{2}\rhonew$. We therefore model the false alarm probability as
$\fap(\rstat) \approx P(\rho/\sqrt{2} | N_t)$. If we have a bank with $N_0$
templates in which the threshold for detection is $\rstat_0$, then to keep the
false alarm probability fixed in a bank with $N_t$ templates, $\rstat$ must
increase to:
\begin{equation}
\label{eqn:rstat_increase}
\rstat^2 = -4\log\left[1 - \left(1-e^{-\rstat_0^2/4}\right)^{N_0/N_t}\right]
\end{equation}
Since templates are not actually independent this model does not give an
accurate absolute value of $\fap(\rhonew)$.  However, the model is adequate
for comparing the \emph{relative} false alarm probabilities of two banks that
cover different size parameter spaces.

\begin{figure}
  \includegraphics[width=\columnwidth]{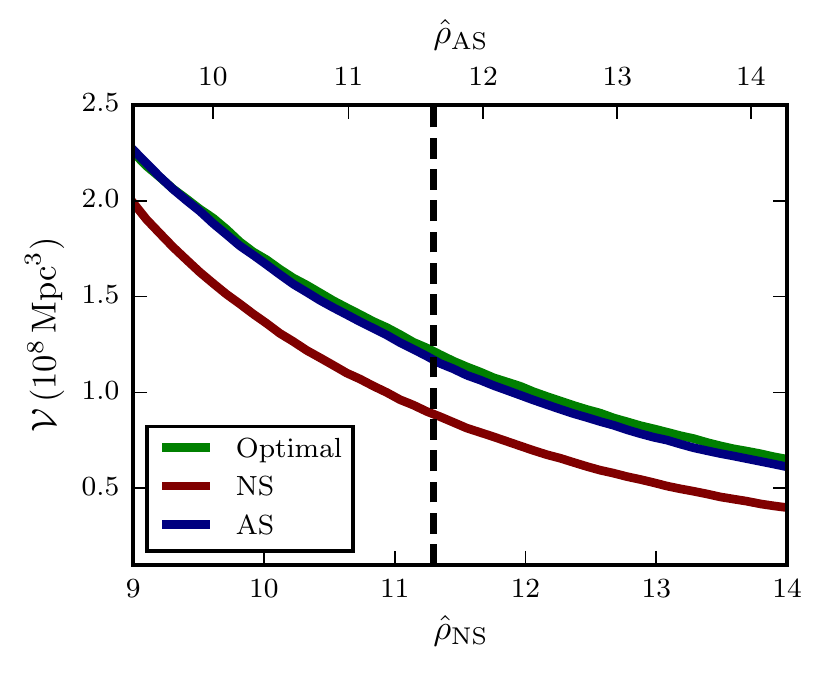}
\caption{Sensitive volume versus re-weighted SNR $\rstat$ of the non-spinning (NS)
and aligned-spin (AS) bank. The bottom axis shows $\rstat$ for the non-spinning bank;
the top axis shows $\rstat$ scaled to account for the increase in false alarm rate of
the aligned-spin bank. The ``Optimal'' volume shows the sensitive volume using the
optimal SNR of each simulated signal as the detection statistic.
The dashed line shows a threshold re-weighted SNR equivalent to a
single-detector SNR threshold of $8$ in the non-spinning search ($\sim 11.3$ for
NS; $\sim 11.7$ for AS).}
\label{fig:gaussian_volume_vs_stat}
\end{figure}

Figure \ref{fig:gaussian_volume_vs_stat} shows the sensitive volume versus
network re-weighted SNR for each bank assuming an astrophysical prior that is
uniform in component masses and spin-magnitude.  The re-weighted SNR of the
aligned-spin bank is offset with respect to the non-spinning bank via
Eq.~\eqref{eqn:rstat_increase} to account for the increase in false alarm
probability.  Even so, we see that the aligned-spin bank is more sensitive than
the non-spinning bank for all thresholds considered in the plot. Also plotted
is the ``optimal'' sensitive volume, which is the sensitivity if every
template matched every signal exactly, and the detection statistic was SNR. We
see that the sensitivity of the aligned-spin bank with re-weighted SNR as the
ranking statistic is close to optimal, as expected from the effectualness
study in the prior section.

A single-detector SNR threshold of $\rho = 8$ is typically assumed to be large
enough to confidently claim a detection. For two detectors, this corresponds to
$\rstat = \sqrt{N_d}\rho \approx 11.3$. We use this as the detection threshold
for the non-spinning bank. By Eq. \eqref{eqn:rstat_increase}, this corresponds
to a threshold of $\approx 11.7$ for the aligned-spin bank. We find that the
aligned-spin bank is $1.30\pm 0.01$ more sensitive than the non-spinning bank
at this threshold (dashed line in Fig.~\ref{fig:gaussian_volume_vs_stat}).

\begin{figure*}
  \begin{center}
  \begin{minipage}{7in}
  \begin{subfigure}{3.375in}
    \includegraphics[width=\textwidth]{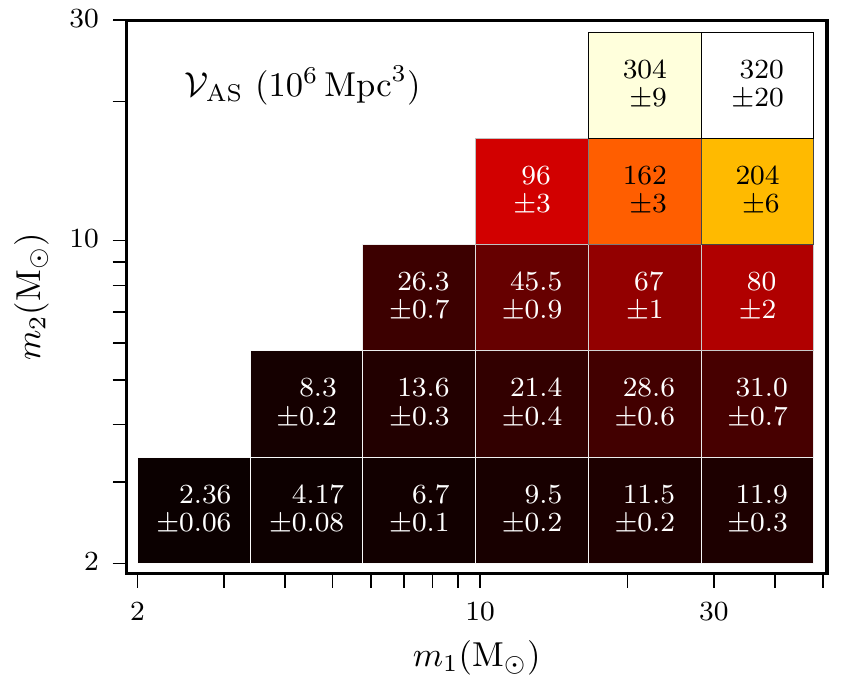}
  \end{subfigure} \hfill
  \begin{subfigure}{3.375in}
    \includegraphics[width=\textwidth]{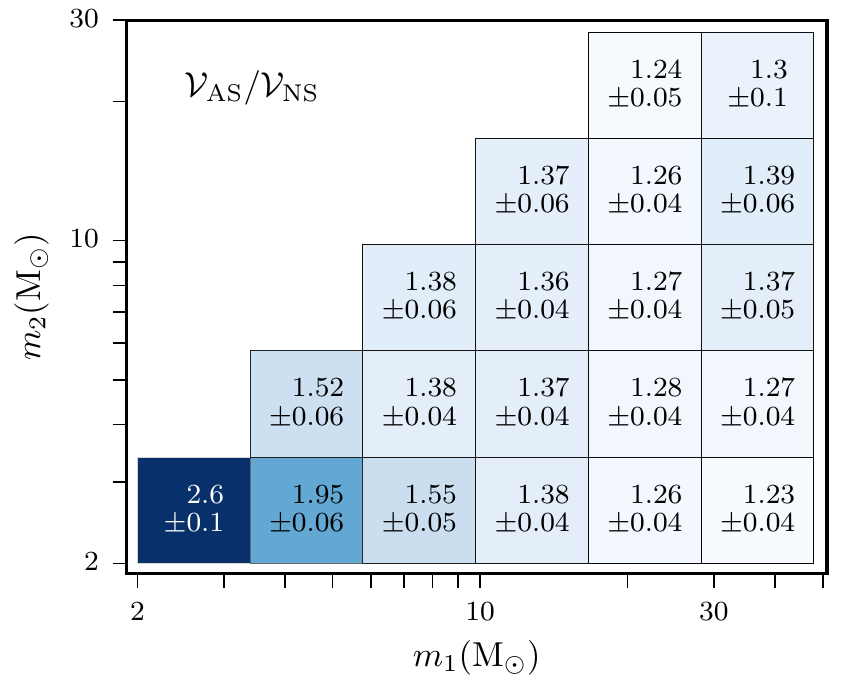}
  \end{subfigure}
  \end{minipage}
  \end{center}
\caption{\textit{Left:} Sensitive volume of the aligned-spin bank
($\vol_{\mathrm{AS}}$) as a function of component mass.
\textit{Right:} Ratio of sensitive volumes of the aligned-spin bank to the
non-spinning bank ($\vol_{\mathrm{NS}}$) as a function of component mass.  The
threshold $\rstat$ used to compute sensitive volumes in these plots is 11.3 for
the non-spinning bank and 11.69 for the aligned-spin bank (dashed line, Fig.
\ref{fig:gaussian_volume_vs_stat}).}
\label{fig:gaussian_volume_over_mass}
\end{figure*}

As stated above, the relative sensitivity of the two banks is strongly dependent on the
astrophysical prior chosen. Indeed, because we have chosen a prior that is
uniform in component masses, the average volume is dominated by high-mass
signals. This can be seen in the left plot of
Fig.~\ref{fig:gaussian_volume_over_mass}, which shows the sensitive volume of
the aligned-spin search sub-divided into bins in component mass. The sensitive
volume of the highest mass tile is over two-orders of magnitude larger than the
lowest-mass tile ($\sim 300\times 10^6\,\mathrm{Mpc}^3$ versus $\sim2\times
10^6\,\mathrm{Mpc}^3$). Taking the ratio of the sensitive volumes of
the aligned-spin bank to the non-spinning bank in each mass bin (right plot
of Fig.~\ref{fig:gaussian_volume_over_mass}), we find that the aligned-spin
bank has a larger gain as we go to lower masses.

\begin{figure*}
  \begin{center}
  \begin{minipage}{7.1in}
  \begin{subfigure}{4in}
    \includegraphics[width=\textwidth]{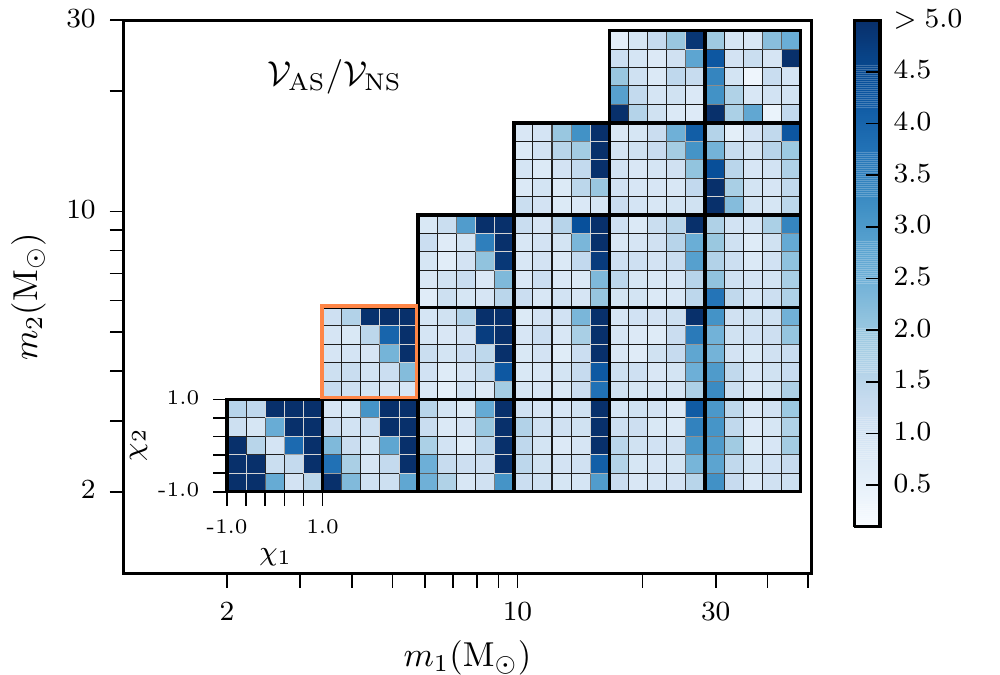}
  \end{subfigure} \hfill
  \begin{subfigure}{3in}
    \includegraphics[width=\textwidth]{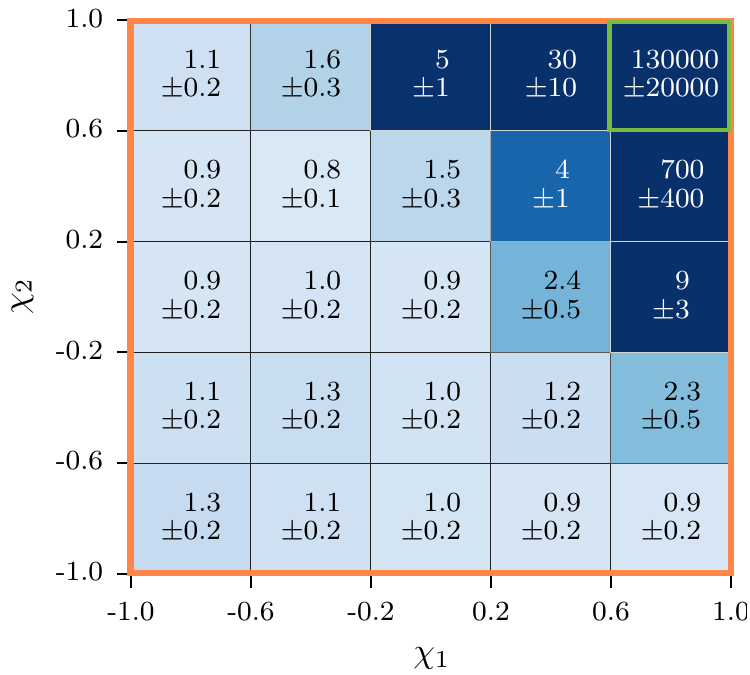}
  \end{subfigure}
  \end{minipage}
  \end{center}
\caption{\textit{Left:} Ratio of sensitive volumes of the aligned-spin bank
($\vol_{\mathrm{AS}}$) to the non-spinning bank ($\vol_{\mathrm{NS}}$) as a
function of component mass and spin. Each mass tile is subdivided into $25$
tiles of the dimensionless spin of each component ($\chi_{1,2}$, inset axes).
Dark blue tiles indicate regions in which the gain is $> 5$.
\textit{Right:} The relative gain as a function of $\chi_1$ and $\chi_2$ for
the mass bin $m_{1,2} \in [3.4, 5.8)\,\Msun$ (highlighted mass tile in the left
plot). The threshold re-weighted SNR used to compute sensitive volume in these
plots is 11.3 for the non-spinning bank and 11.7 for the aligned-spin bank
(dashed line, Fig.  \ref{fig:gaussian_volume_vs_stat}).}
\label{fig:gaussian_gain}
\end{figure*}

In Fig.~\ref{fig:gaussian_gain} (left) we further sub-divide each mass tile into
$25$ bins in $\chi_1$ and $\chi_2$, with the color bar indicating the relative
sensitivity of the two banks. In the lowest-mass tile (bottom left corner), we
see that the gain in sensitivity can be $> 5$ when both spins are either
$<-0.2$ (anti-aligned) or $>0.2$.  As we move to higher mass tiles, however,
the gain in the anti-aligned region approaches one, particularly for tiles that
are close to the equal-mass line.  This is likely due to the degeneracy between
spin and the symmetric mass ratio $\eta = (m_1m_2)/M_{\mathrm{total}}^2$. Namely, at lower
masses for which the inspiral dominates the waveform, negative (positive) spins
are degenerate with smaller (larger) $\eta$ \cite{Ohme:2013nsa}. Signals with
negative spins can therefore be recovered by the non-spinning bank by templates
with smaller $\eta$, which corresponds to larger mass ratio. In the lowest-mass
corner of the bank, the range in mass ratio is limited by our minimum component
mass boundary of $2\,\Msun$. We expect that if we extended the component mass
to lower masses, the resulting larger range in mass ratio would mitigate the
non-spinning bank's loss in sensitivity to anti-aligned spins. Indeed, in the
full search a minimum component mass of $1\,\Msun$ is used to incorporate BNS
and NSBH systems \cite{TheLIGOScientific:2016qqj}.

While the non-spinning bank's loss in sensitivity to low-mass, anti-aligned
signals can be mitigated by extending the bank's boundaries to lower $\eta$,
the same is not true for positive aligned-spins. This is because the symmetric
mass ratio reaches a maximum at $0.25$, which corresponds to equal-mass
systems. To extend the non-spinning bank to larger $\eta$ would require
non-physical component masses.  This is evident in
Fig.~\ref{fig:gaussian_gain}: the gain in sensitivity remains $\gtrsim 5$ for
component spins $\gtrsim 0.2$ and component masses $\lesssim 20\,\Msun$ even as
the range in mass ratios spanned by the bank increases. This gain can be more
than an order-of-magnitude when the component spins are $>0.6$. Indeed, if we
zoom in on one of these low-mass tiles, $m_{1,2} \in [3.4, 5.8)\,\Msun$
(highlighted box in the left plot), we find that aligned-spin bank is up to
$O(10^5)$ times more sensitive than the non-spinning bank for $\chi_{1,2} >
0.6$ (highlighted tile in the right plot of Fig.~\ref{fig:gaussian_gain}).
This is much larger than what is expected from SNR loss alone.  The
effectualness of the non-spinning bank is between $\sim0.7$ and $\sim0.6$ for
this region of parameter space, indicating an SNR loss of $30$ -- $40\,\%$.
That would translate to a loss in sensitive volume of $66$ -- $78\,\%$.

\begin{figure}
  \begin{center}
  \begin{subfigure}{\columnwidth}
    \includegraphics[width=\textwidth]{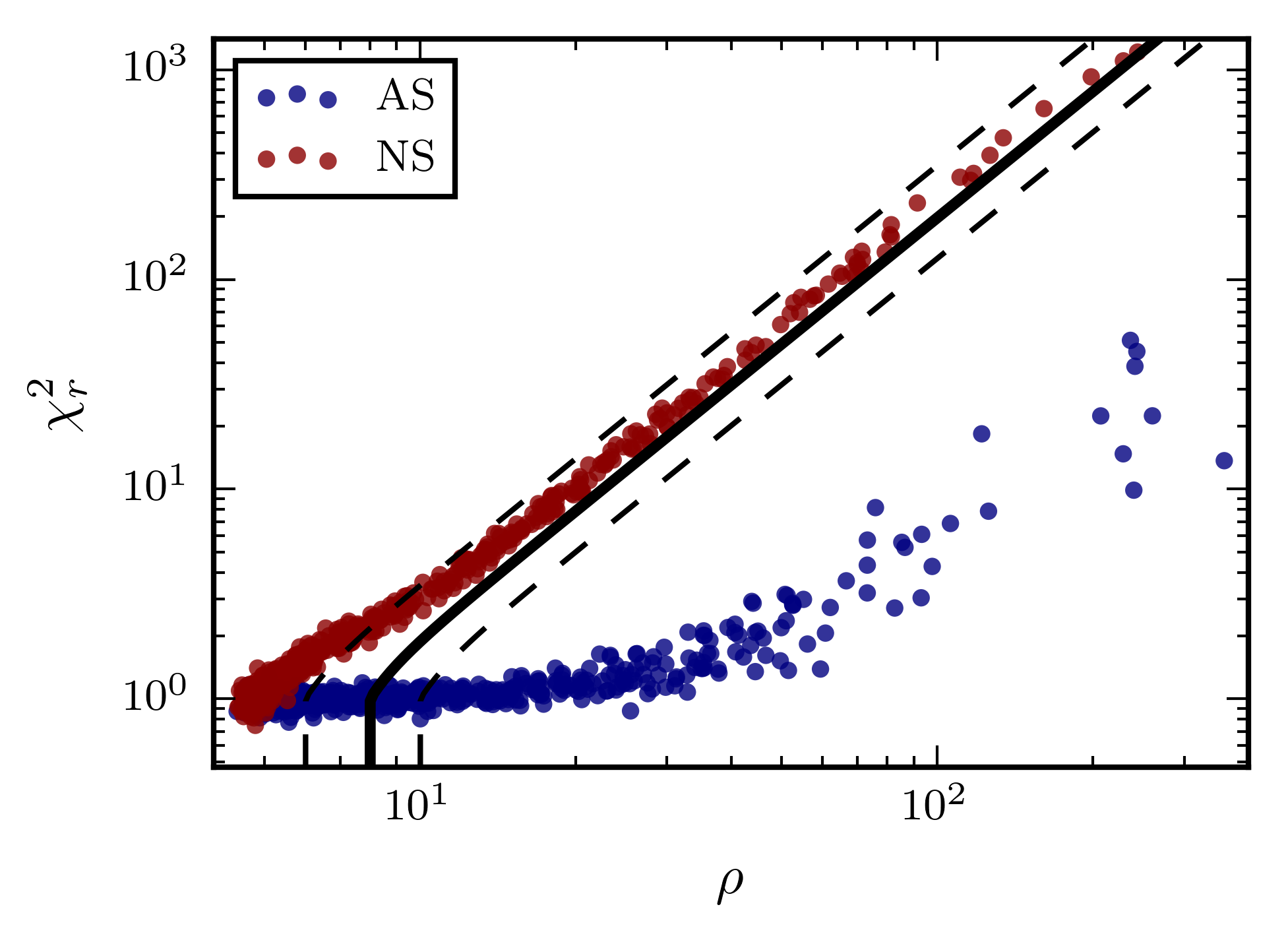}
  \end{subfigure}
  \\
  \begin{subfigure}{\columnwidth}
    \includegraphics[width=\textwidth]{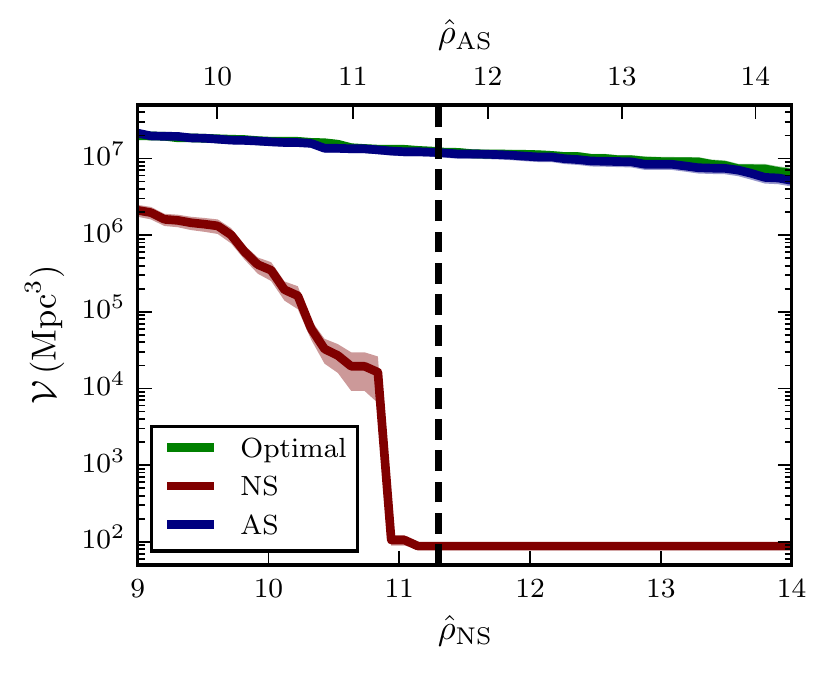}
  \end{subfigure}
  \end{center}
\caption{Single-detector SNR ($\rho$) versus reduced $\chi^2$ (\emph{top}),
and sensitive volume versus threshold $\rstat$ (\emph{bottom}) for signals with
$m_{1,2} \in [3.4,5.8)\,\Msun$ and $\chi_{1,2} \geq 0.6$ (the highlighted tile in
the bottom plot of Fig.~\ref{fig:gaussian_gain}).
The dashed lines in the top plot show lines of constant single-detector
re-weighted SNR; the thick solid line shows a re-weighted SNR $=8$, which is
the threshold we used for the non-spinning bank.}
\label{fig:gaussian_gain_reason}
\end{figure}

The reason for the large increase in sensitivity can be seen in the top plot of
Fig.~\ref{fig:gaussian_gain_reason}, which shows the reduced $\chi^2$ versus
SNR in a single detector for this region of parameter space. We see that the
$\chi_r^2$ of the signals when recovered by the non-spinning bank are
quite large. In fact, the $\chi_r^2$ statistic asymptotes such that the re-weighted
SNR of the signals is always $< 8$ (solid black line in the figure). Thus, even
when the optimal SNR of a signal is $400$ (and the recovered SNR by the
non-spinning bank is $200$), the re-weighted SNR is still less than 8, which is
the threshold for detection. This is strongly dependent on the threshold for
detection: if the single-detector threshold was 6 (top dashed-line), these
high-SNR events would be detected.  This can be seen in the bottom plot of
Fig.~\ref{fig:gaussian_gain_reason}, which shows the volume versus threshold
network re-weighted SNR for just this region of parameter space. We see that as
the threshold decreases below 11.3 (corresponding to a single-detector SNR $=
8$), the sensitivity of the non-spinning bank rapidly improves. For instance,
at a threshold of $\rhonew = 8$ (which corresponds to a single-detector SNR
$\sim 5.6$), the gain is $\sim 3$, closer to that predicted by the loss in SNR.

The non-spinning sensitivity levels off at $\rhonew_{\mathrm{NS}} \sim 11$ in
the bottom plot of Fig.~\ref{fig:gaussian_gain_reason} due to the minimum
distance bound we used for the simulated signals. If we had chosen smaller
bounds, the $\vol_{\mathrm{NS}}$ would continue to drop, indicating that the
non-spinning bank has zero sensitivity to these signals. We find similar
characteristics for tiles in which the component masses are $< 17\,\Msun$,
$\chi_1 \geq 0.6$, and $\chi_2 \geq -0.2$.

By only filtering the most effectual template, we have made the assumption that
$\left<\mathcal{V}(\rstat)\right> \approx \mathcal{V}(\left<\rstat\right>)$.
Figure \ref{fig:gaussian_gain_reason} indicates that we are in a regime where
small changes in $\chi_r^2$ have large effects on the sensitive volume. We expect
that the approximation breaks down in this regime. The $\chi_r^2$ values of
signals will fluctuate about the mean in different realizations of noise, which
will cause small fluctuations in $\rstat$.  For these low-mass, high-spin areas
of parameter space, we expect these small fluctuations to be enough to occasionally
bring $\rstat$ above threshold. Thus we do not expect the sensitive volume of
the non-spinning bank to be exactly zero. In the next section we filter these low-mass,
high-spin signals with the full template bank and pipeline to get a better
estimate of the gain between the non-spinning bank and the aligned-spin bank
in these regions of parameter space.

\subsection{Recolored noise}
\label{ssec:recolored_comparison}

To get a better estimate of the gain in sensitivity we can expect between the
aligned-spin and non-spinning banks, we add simulated signals to S6 data
recolored to resemble early advanced LIGO data. We
analyze that data using the search algorithm described in Refs.~\cite{Canton:2014ena,Usman:2015kfa}.
That is, we filter
each signal with all templates, find coincidence, maximize over the bank using
network re-weighted SNR, and estimate the background to find $\fap(\rstat)$.
Since this is computationally expensive to perform on all $500\,000$ simulated
signals, we limit this study to areas of parameter space where the gain in
sensitive volume was $\gg 5$ in the previous section. Namely, we restrict to
signals with $m_{1,2} < 17$, $\chi_1 > 0.6$, and $\chi_2 > -0.2$, of which there
are $\sim 10\,000$.

As discussed in the last section, due to the presence of glitches, the
distribution of $\rstat$ is not the same for all templates in real noise, as it
is in Gaussian noise. Results are typically binned by some parameter
\cite{Colaboration:2011np, Aasi:2012rja} when estimating false alarm probability for this
reason.  For example, in Ref. \cite{Colaboration:2011np}, three bins in chirp mass were
used.  We do not try to do any binning here. This means that our results may
not be as optimal, but we expect such binning to have a small effect on the
relative gain in sensitivity between the non-spinning and aligned-spin bank.

\begin{figure}
  \includegraphics[width=\columnwidth]{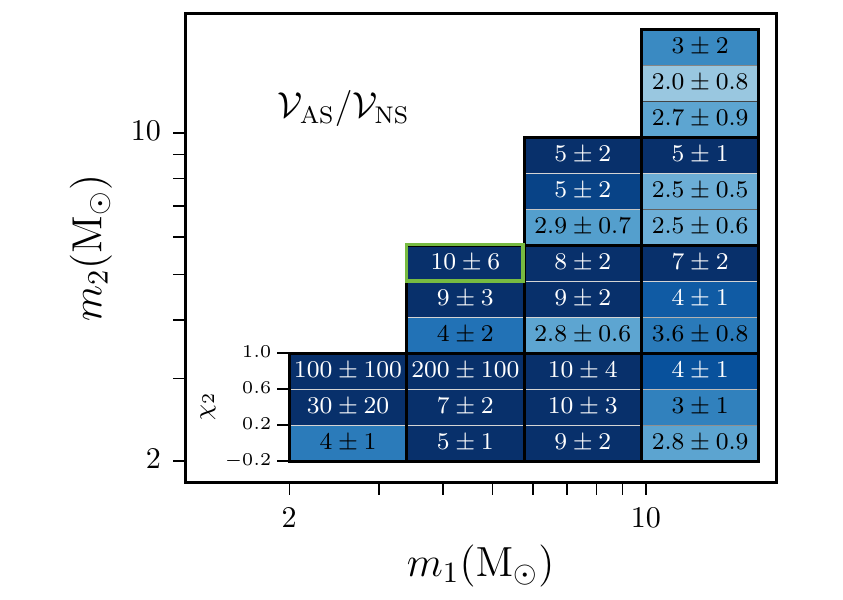}
\caption{Ratio of sensitive volumes of the aligned-spin bank
($\vol_{\mathrm{AS}}$) to the non-spinning bank ($\vol_{\mathrm{NS}}$) as a
function of component mass and spin in the recolored noise. Tiles are only
shown for the simulated signals that were selected for this test (
$\chi_1 > 0.6,~\chi_2 \in [-0.2,0.99), m_{1,2} < 16.6\,\Msun$). The highlighted
tile corresponds to the same region of parameter space as the highlighted tile
in the right plot of Fig.~\ref{fig:gaussian_gain}.}
\label{fig:recolored_gain}
\end{figure}

\begin{figure}
  \begin{center}
  \begin{subfigure}{\columnwidth}
    \includegraphics[width=\textwidth]{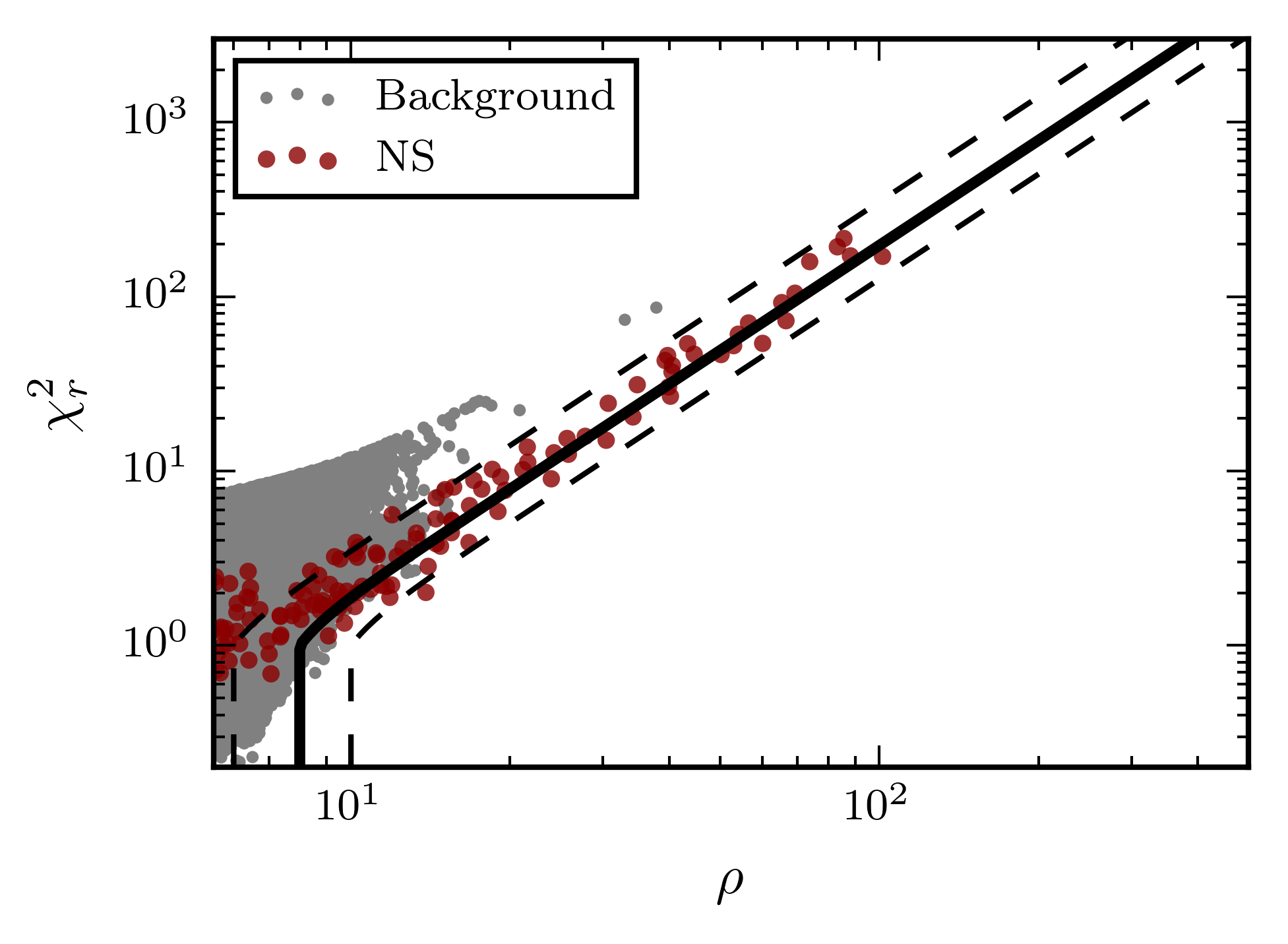}
  \end{subfigure}
  \\
  \begin{subfigure}{\columnwidth}
    \includegraphics[width=\textwidth]{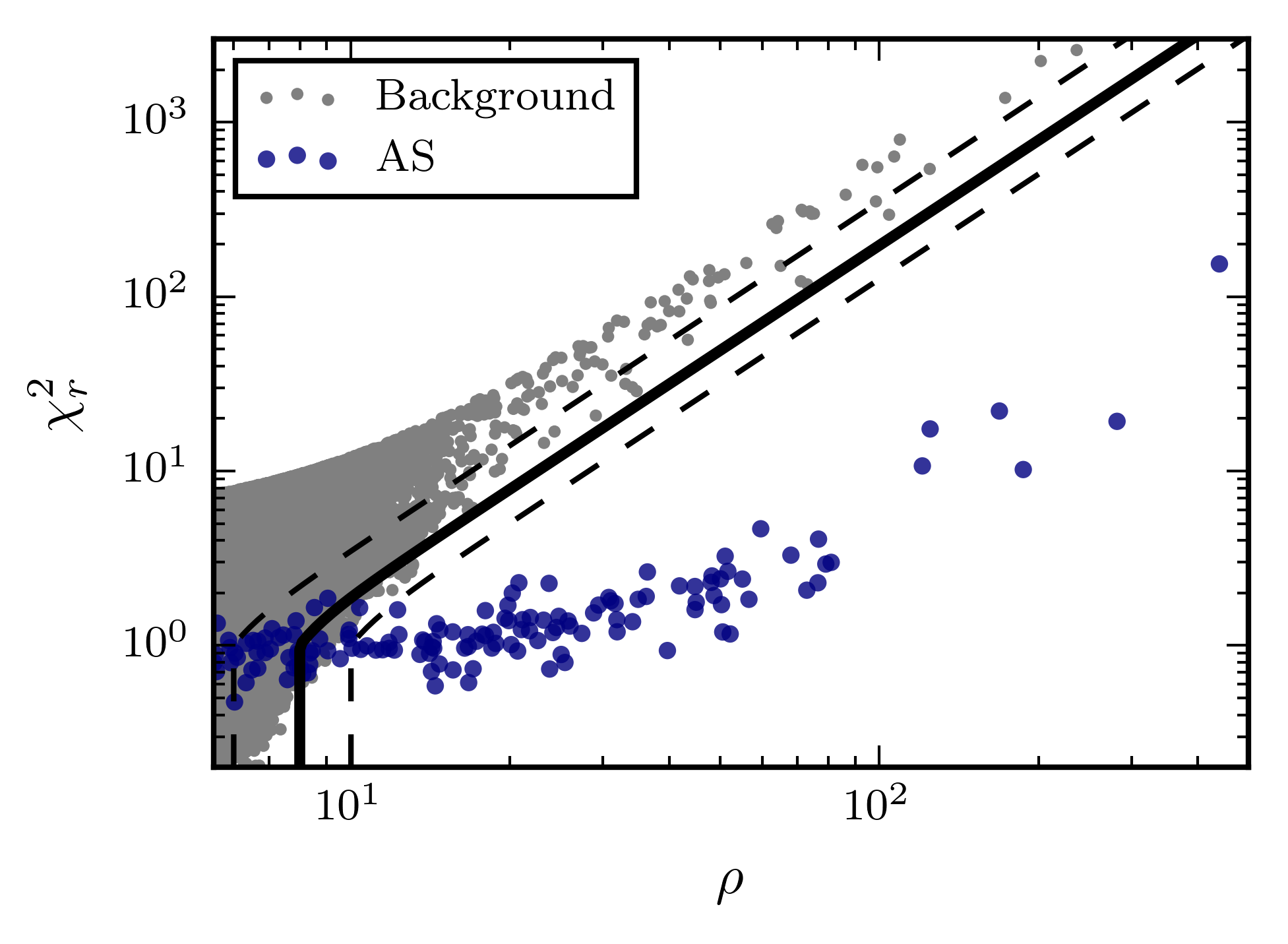}
  \end{subfigure}
  \end{center}
\caption{Single-detector SNR ($\rho$) versus reduced $\chi^2$ for the non-spinning (\emph{top})
and aligned-spin (\emph{bottom}) banks in recolored noise for signals with
$m_{1,2} \in [3.4,5.8)\,\Msun$ and $\chi_{1,2} \geq 0.6$ (the highlighted tile in
the right plot of Fig.~\ref{fig:gaussian_gain}) Also shown are the false alarms from
each search (gray dots).}
\label{fig:recolored_snrchi}
\end{figure}

Figure \ref{fig:recolored_gain} shows the relative gain between the
non-spinning bank and the aligned-spin bank in the recolored noise. As
expected, the gain is not as large as we found in the previous section. For
example, focusing on the same region of parameter space that we highlighted in
the prior section ($m_{1,2} \in [3.4,5.8)\,\Msun$ and $\chi_{1,2} \geq 0.6$)
the gain is $\exTileRecoloredGain$ (highlighted tile in Fig.
\ref{fig:recolored_gain}). The reason for this can be seen in
Fig.~\ref{fig:recolored_snrchi}, which shows reduced $\chi^2$ versus SNR in a
single detector for these signals when recovered by the non-spinning bank
(top) and the aligned-spin bank (bottom) in the recolored noise. We find
similar behavior as in the Gaussian noise results
(Fig.~\ref{fig:gaussian_gain_reason}): the $\chi^2$ of the signals when
recovered by the non-spinning bank is large, causing the signals to asymptote
to lines of constant re-weighted SNR.  Even so, the variance of the reduced
$\chi^2$ causes some signals to have re-weighted SNRs greater than the
threshold of 8.  The result is the non-spinning bank does recover at least some
of the signals, yielding a non-zero sensitive volume. However, the sensitive
volume of the aligned-spin bank in these areas of parameter space can still be
an order-of-magnitude or larger than the non-spinning bank, which is larger
than one might expect from SNR loss alone.

Triggers caused by noise are also shown in Fig. \ref{fig:recolored_snrchi}
(gray dots in both plots). These triggers form the background with which false
alarm probability is measured. To improve the sensitivity of the non-spinning
bank, one might consider changing the exact form of the $\chi^2$ re-weighting
of the SNR used in Eq.~\eqref{eqn:newsnr}.  However, this would be difficult to
do without also promoting noise triggers to higher significance. The large
mismatch between the non-spinning templates and the spinning signals makes it
difficult to separate glitches from signals. Contrast this to the aligned-spin
bank.  Although the number of noise triggers has increased, there continues to
be good separation between the aligned-spin signals and noise. We conclude that
the aligned-spin bank can be implemented using the same form of the re-weighted
SNR given in Eq.~\eqref{eqn:newsnr}.

\section{Conclusions}
\label{sec:conclusions}

We have demonstrated here a complete method for conducting a search for GWs
from BBH using an aligned-spin template bank in advanced LIGO. We have covered
the parameter space combining two previously proposed methods for template
placement: geometric and stochastic. We have shown that combining these
methods yields a more effectual template bank than the geometric method alone,
while also using $\sim5\%$ fewer templates than the stochastic method alone.
We expect that the savings will only increase as the lower-frequency
performance of the LIGO detectors improves in future observing runs.

Applying the template bank to an analysis of simulated advanced LIGO data, we
have characterized the improvement in sensitivity of the pipeline towards
aligned-spin signals. We have found that the aligned-spin bank is significantly
more sensitive than the non-spinning bank to signals with $\chi_{1} \gtrsim
0.6$ and component masses $m_{1,2} \lesssim 20\,\Msun$. From mismatch alone, we
would expect the aligned-spin bank to have a sensitive volume that is $20$ to
$30\%$ larger than the non-spinning bank in this region of parameter space.
However, when the effects of $\chi^2$ re-weighting of SNR are included, we find
that the aligned-spin bank can be one to two orders-of-magnitude more sensitive
to these signals. Although less pronounced, the aligned-spin bank is also more
sensitive to higher-mass systems with non-zero spins, with gains of $30-500\%$.

Due to the increase in false alarm rate, adding the aligned spin templates does
reduce the sensitivity to non-spinning systems by $\sim10\%$. However, this
would only lead to a loss in detection rate if nearly all systems in the
universe were non-spinning. Based on observations of X-ray binaries
~\cite{McClintock:2006xd, Gou:2011nq, Gou:2010qq, Gou:2009ks, Miller:2009cw,
Valtonen:2010xf, Martocchia:2002wt} and population synthesis
models~\cite{Belczynski:2007xg} we expect many systems to have spin. The
significant gain in sensitivity to spinning systems therefore compensates
for the relatively small loss in sensitivity to non-spinning systems.

Although we only considered BBHs with total mass $\leq 50\,\Msun$ in this
study, the template-placement methods discussed here can be applied to a larger
range of masses. We expect the gain in sensitivity to become less dramatic for
signals with total masses $> 50\,\Msun$, even if $\chi_1 \gtrsim 0.6$. This is
because $\chi^2$ becomes less effective at higher masses, in which the
bandwidth of templates is short.

In order to arrive at these results we used the same tuning as was used in the
search for CBCs in the sixth science run of initial LIGO
\cite{Colaboration:2011np}. Namely, we used $16$ bins when computing $\chi^2$,
and we used Eq.~\eqref{eqn:newsnr} to re-weight SNR. Due to the large $\chi^2$
values, we found that the sensitivity of the non-spinning bank can vary
dramatically with small changes in the threshold $\rstat$ used to detect
signals. Although another choice of $\chi^2$ bins and re-weighting is possible,
it would be difficult to improve the sensitivity of the non-spinning bank
without decreasing the ability of the search to separate signals from glitches.
This is due to the large mismatches involved between non-spinning templates and
spinning signals. The simplest, safest approach is to simply use aligned-spin
templates in the search.

An aligned-spin template bank was used in the modeled search that found
GW150915 \cite{Abbott:2016blz,TheLIGOScientific:2016qqj} that used the
combined geometric and stochastic placement method discussed here. While
GW150914 did not have very large spin --- its effective spin was estimated
to be $-0.06\pm0.18$~\cite{TheLIGOScientific:2016wfe} --- future detections
may yield black holes with significant spin.

\section{Acknowledgements}

We are grateful to Sukanta Bose, Thomas Dent, Chad Hanna, Andrew Lundgren, Alex
Nielsen, and Alex Nitz for providing insightful discussions. The authors would
also like to thank the Max Planck Gesellschaft for support. All calculations
were carried out on the ATLAS and VULCAN clusters at the Albert Einstein
Institute in Hannover and Potsdam, Germany, respectively.

\appendix

\section{Derivation of sensitive volume}

The integrand of Eq. \eqref{eqn:sensitive_v} is a complicated function of the integration variables;
the efficiency, for example, depends on the characteristics of the data and how well the ranking
statistic $\rstat$ separates noise from signal, which is not known \emph{a priori}. We therefore find
the sensitive volume by Monte Carlo integration, as follows.

First, note that the efficiency function serves to determine what fraction of the total
number of signals are detected. For example, if the universe contained $N$ signals within
volume $V_{\max}$ (i.e., $q$ was a series of $N$ delta functions), $n$ of which are
detected by a pipeline at some threshold $\rstat$, then the sensitive volume would be $n V_{\max}/N$.
Assume then that we have $N$ random simulated signals drawn from the same distribution as $q$, which
we filter through the pipeline to acquire a ranking stat value $\rstat_i$ for each. We can replace
$\epsilon$ with a step function $\Theta$ such that $\Theta(\rstat) = 1$ if $\rstat \geq \rstat_i$, and
0 otherwise. The sensitive volume is then:
\begin{equation}
\label{eqn:mcV}
\vol(\rstat) \approx V_{\max} \frac{1}{N} \sum_{i}^{N} \Theta(\rstat) = V_{\max} \left<\Theta(\rstat)\right>,
\end{equation}
The error in this estimate is given by the square root of the variance:
\begin{equation}
\label{eqn:mcVerr}
\delta \vol = V_{\max} \sqrt{\frac{\left<\Theta^2\right> - \left<\Theta\right>^2}{N}}.
\end{equation}
%

As discussed in the main text, for BBHs with total masses $\leq 50\,\Msun$, the
LIGO detectors are sensitive out to a maximum distance of $\sim1\,\mathrm{Gpc}$
assuming the early advanced LIGO PSD. Over these distances we assume an
isotropic distribution of signals. Thus for Eq. \eqref{eqn:mcV}
to be correct, we have to draw our simulated signals from a spatial
distribution that is uniform in volume.  However, due to the antenna pattern of
the detectors, we have found that using a distribution uniform in volume causes
most of the signals to be too weak to be detected by the pipeline. This leads
to a large variance in the volume estimate. A more efficient approach is to do
importance sampling, i.e., to draw the simulated signals from a distribution
$q'$ such that the majority of the signals closely straddle the boundary
between being detected (\emph{found}) and not being detected (\emph{missed}).

In general, any Monte Carlo integral of a function $f$ over some volume $V$ can be written
\cite{NumMethods:2007}:
\begin{equation}
\label{eqn:importancesamp1}
I = \int\limits_{V} f(\spatialV) q(\spatialV) \mathrm{d}\spatialV
  = \int\limits_{V} g(\spatialV) q'(\spatialV)\mathrm{d}\spatialV,
\end{equation}
where $g(\spatialV) = f(\spatialV)q(\spatialV)/q'(\spatialV)$. Here, $q$ is the distribution
that is uniform in $V$ and $q'$ is any other distribution from which we would prefer to draw
samples. In order to approximate this as an average over $N$ random points sampled from $q'$ we need to
transform to the coordinate system $\spatialV'$ in which $q'$ is uniform. Let $s^{-1}$ be the
transformation from $\spatialV'$ to $\spatialV$, such that $s(\spatialV') = \spatialV$, and $|\mathbf{J}|$
be the Jacobian determinant of the transformation (i.e., $J_{ij} = \partial x'_i / \partial x_j$).
The integral is then the weighted average:
\begin{align}
\label{eqn:importancesamp2}
  I &= \int g\left(s[\spatialV']\right) q\left(s[\spatialV']\right)|\mathbf{J}| \mathrm{d}\spatialV' \nonumber \\
    &\approx \sum_{i=1}^{N} \tilde{w}_i f\left(s[\spatialV'_i]\right) \equiv \left<f\right>_{w}
\end{align}
where:
\begin{equation}
\label{eqn:importancesamp_weight}
\tilde{w}_i = \frac{w_i}{\sum_{j=1}^{N} w_j}; \quad
w_i = \frac{q\left(s[\spatialV']\right) |\mathbf{J}|}{q'\left(s[\spatialV']\right)}.
\end{equation}
This has error $\sqrt{(\left<f^2\right>_w - \left<f\right>_w^2)/N}$ \cite{1995AtmEn..29.1185G}.

We apply this to the spatial part of the sensitive volume integral as follows.
First, we assume that within some minimum distance $r_{\min}$ nearly all
signals will be detected by the pipeline at our detection threshold $\threshrstat$.
Likewise, due to noise, we know that all signals beyond some maximum distance
$r_{\max}$ will not be detected by the pipeline. We determine these bounds
based on the optimal network SNR. The optimal network SNR $\rho_{\mathrm{opt}}$
of a signal with some intrinsic and extrinsic parameters $\allparams^*$ is
related to its physical distance $r$ via:
\begin{equation}
\label{eqn:rToSNR}
r = \frac{\sqrt{\sum_{i} \left<h_i(\allparams^*), h_i(\allparams^*)\right>}}{\rho_{\mathrm{opt}}},
\end{equation}
where $h_i$ is the strain caused by the signal at a fiducial distance of
$1\,$Mpc in the $i$th detector, and the sum is over the number of detectors. If
$\rstat = \rho_{\mathrm{opt}}$, then we know that the signal would be missed
(found) at distances greater (less) than the distance corresponding to
$\rho_{\mathrm{opt}} = \threshrstat$. However, due to the mismatch between
signal and template, $\chi^2$ re-weighting, and the presence of noise, $\rstat$
is not exactly equal to $\rho_{\mathrm{opt}}$. We therefore choose a maximum
and minimum $\rho_{\mathrm{opt}}$ that we are confident bounds
$\threshrstat$.  We then obtain a minimum and maximum distance bound for each
simulated signal via Eq.  \eqref{eqn:rToSNR}.  In this study, our detection
threshold is $\rstat^* = 8 \sqrt{2}$; we conservatively choose a maximum
(minimum) $\rho_{\mathrm{opt}}/\sqrt{2}$ of 400 (4) to obtain the distance
bounds. Note that this means that $r_{\min}$ and $r_{\max}$ depend on the
intrinsic and extrinsic parameters of the signal.

Given the distance bounds, we next choose a distribution to draw the distances
from, $q'$.  This choice is informed by our assumed distribution of signals
$q$.  As discussed in the main text, for this study we assume an isotropic
distribution of signals in the universe; i.e., we assume that $q$ is
independent of $\spatialV$. Given this distribution of signals and our choice
of distance bounds, we have found that using a distribution $q'$ uniform in the
solid angle $\Omega$ and uniform in distance yields volume estimates with
reasonably small variance. Thus for $r \in [r_{min}(\allparams), r_{\max}(\allparams))$,
\begin{equation*}
q'(\allparams, r) = [r_{\max}(\allparams) - r_{\min}(\allparams)]^{-1} \equiv \Delta r(\allparams).
\end{equation*}
Since $q'$ is uniform in $r$ and $\Omega$, the Jacobian determinant $|\mathbf{J}| = r^2$.
With these choices, the sensitive volume is:
\begin{align}
\label{eqn:weightedMCV}
\vol(\rstat) &= \int \mathrm{d}\Omega \int \mathrm{d}\allparams q(\allparams) \left[\int_{0}^{r_{\min}(\allparams)} r^2\mathrm{d}r \right. \nonumber \\
    &\left. \quad \quad + \int_{r_{\min}(\allparams)}^{r_{\max}(\allparams)} \Theta(\rstat; r, \allparams) \Delta r(\allparams) r^2 \mathrm{d}r \right] \nonumber \\
 &\approx 4\pi \frac{1}{N} \sum_{i=1}^{N} \left[\frac{1}{3}r_{\min,i}^3 + \Theta_i(\rstat) \Delta r_i  r^2_i \right].
\end{align}
The sum is over the simulated signals, the intrinsic and extrinsic parameters of which are drawn from $q(\allparams)$,
and sky locations drawn uniform in the solid angle $\Omega$.

Equation \eqref{eqn:weightedMCV} assumes that the distribution of signals in
the universe over $\allparams$ is the same as the distribution of the simulated
signals. That is, we have applied a weight of $1/N$ to all of the
injections. We can also use Eqs. \eqref{eqn:importancesamp1} -
\eqref{eqn:importancesamp_weight} to weight the simulated distribution to any
other astrophysical distribution that we believe to be plausible. This allows
us to test different distributions (about which we are uncertain) while still
using the same set of simulated signals. This saves on computational costs,
though it does result in a larger variance in the sensitive volume estimate.

The most compuationally expensive step in the Monte Carlo simulation is
filtering the simulated signals to find the largest SNR over the bank. We
already have these results from the effectualness studies in Sec.
\ref{sec:bank_construction}. However, in those studies we use a simulation
distribution $q'(\allparams)$ that is $\log$ in the component masses, while our
assumed astrophysical distribution $q(\allparams)$ is uniform in the component
masses. To account for this, we reverse the prescription we used above; i.e.,
we determine the weight needed to convert from the simulated distribution
$q'(\allparams)$ to the assumed distribution $q(\allparams)$.  The Jacobian
determinant for this transformation is $m_1 m_2$. The sensitive volume estimate
is thus:
\begin{equation}
\label{eqn:logweightedMCV}
\vol(\rstat) \approx 4\pi \sum_{i=1}^{N} \tilde{w}_i\left[\frac{1}{3}r^3_{\min,i} + \Theta_i(\rstat)\Delta r_i r^2_i\right],
\end{equation}
where
\begin{equation}
\label{eqn:normed_weights}
\tilde{w}_i = \frac{m_{1i}m_{2i}}{\sum_{j=1}^{N} m_{1j}m_{2j}},
\end{equation}
are the normalized weights needed to convert between the mass distributions.

\bibliography{references}

\end{document}